# Discharge structure of $Ar/Cl_2$ inductively coupled plasma: A cyclic study of discharge conditions at fixed power

[a]Shu-Xia Zhao[1] and An-Qi Tang[2]

1. *Key Laboratory of Material Modification by Laser, Ion, and Electron Beams (Ministry of Education), School of Physics, Dalian University of Technology, 116024, Dalian, China*
2. *China Key System & Integrated Circuit Co., Ltd, 214026, Wuxi, China*

[a] *Correspondence: zhaonie@dlut.edu.cn*

## Abstract

Discharge structure refers to the morphology of different plasma quantities, such as electron temperature, reaction rate, plasma potential, mass flux, net charge and species density, which are determined by plasma transport mechanism and chemical processes. This morphology is rather difficult to refine in complex electronegative plasma for it is contained in a multiple-physics-field coupled system. Regarding this difficulty, the combination of self-consistent simulation, theory analysis and experimental diagnostic of this system is needed. In the scope of present article, the fluid simulation and analytic theory are utilized to investigate the $Ar/Cl_2$ inductive plasma, via a cyclic tuning of discharge pressure and feedstock gas content parameters when fixing the power. Classic discharge structure (*e.g.*, delta, parabola, flat-top, and hollow) and specific discharge mechanisms (*e.g.*, self-coagulation, physics coagulation, de-coagulation, grouping behavior and the ambi-polar diffusion of triple species) are revealed. Besides, the two types of discharge stratification, *i.e.*, space and species, are presented. Many non-neutralities are generated during the transport process of electronegative plasma when it gives rise to the discharge structure, which are summed and analyzed for better understanding the plasma.

## I. Introduction

Gaseous discharge structure of laboratory plasma is very important issue, which has been investigated by many researchers. At initial, the spatially resolved plasma density, electron temperature and electric potential of argon plasma were measured by Langmuir probe [1] and validated by fluid or hybrid models [2,3]. For reactive plasmas, Han *et al* used the same method to investigate the electron density profile of $Ar/O_2$ inductively couple plasma [4]. It is noted that these comparative studies cannot reveal the transport scheme of plasma and neither explains the profile characteristics. Besides for the experiments and self-consistent simulations, early researchers still paid attention on establishing analytic theories for better understanding the discharge structure. They started from the electropositive plasma, and generated many useful conclusions by using drift-diffusion flux and continuity equation, together with the simple chemistry. Parabola, trigonometric, and Bessel distributions of plasma density were given with another approximation, ambi-polar diffusion [5]. These researchers later extended the above analytic method content and shifted it into the electronegative plasma field. They introduced the anion Boltzmann, and sequentially built the parabola (suitable for electronegative plasma), ellipse and flat-top models [6,7]. Other academicians utilized the similar analytic theory to predict the species-stratification, parabola and uniform profiles, by means of their own definitions of different models [8]. Besides, there are still many other phenomena discovered, such as the space-stratification [9], double layer [10], ions acoustics [11], and internal sheath [12] etc. These are more about the microscopic characteristic of discharge structure. These analytic works guided well the experimental activities of that time, for many diagnostics of electronegative plasma compared their ions density with parabola and flat-top models [13-15]. Relatively, the self-consistent simulation works are not related to the above analytic works. Many researchers of simulation field even are not familiar with these theories. This leads to difficulties on analyzing the simulated profiles, *i.e.*, discharge structure. To us, not only the densities are important, but also other quantities, such as charge density, mass flux, chemical source, species velocity and so on. The present work of this article is aimed at solving this problem, by taking the $Ar/Cl_2$ inductive plasma as one example at a cyclic tuning of discharge parameter. Together with the previous correlative works of $Ar/SF_6$ and $Ar/O_2$ inductive plasmas [16,17] where the self-coagulation theory was constructed, we plan to build up the elaborate discharge structure of electronegative plasmas, first by means of fluid model simulation and analytic theories. Then validated by experiments and amend by kinetic effects.

## II. Fluid model and $Ar/Cl_2$ chemistry

As mentioned, in this work the fluid model and analytic theory are used to analyze the $Ar/Cl_2$ inductive plasma discharge structure. The fluid model is described in Ref. [16], together with the analytic theories, including the old ones, such as flat-top model and the new self-coagulation. Here in this section, only the chemistry and surface kinetics are introduced, which are listed in Tabs. 1 and 2.

Table 1. Chemical reaction set considered in the model

| No. | Reaction | Rate coefficient[a] | Threshold (eV) | Ref. |
|---|---|---|---|---|
| 1 | $e+Ar \rightarrow e+Ar$ | Cross Section | 0 | 18 |
| 2 | $e+Ar \rightarrow e+Ar^m$ | Cross Section | 11.6 | 18 |
| 3 | $e+Ar^m \rightarrow e+Ar$ | Cross Section | -11.6 | 18 |
| 4 | $e+Ar \rightarrow 2e+Ar^+$ | Cross Section | 15.76 | 18 |
| 5 | $e+Ar^m \rightarrow 2e+Ar^+$ | Cross Section | 4.43 | 18 |
| 6 | $e+Cl_2 \rightarrow e+Cl_2$ | Cross Section | 0 | 18 |
| 7 | $e+Cl \rightarrow e+Cl$ | Cross Section | 0 | 18 |
| 8 | $e+Cl_2 \rightarrow e+Cl_2(v1)$ | $4.35\times10^{-10}T_e^{-1.48}\exp(-0.76/T_e)$ | 0.07 | 19,20 |
| 9 | $e+Cl_2 \rightarrow e+Cl_2(v2)$ | $8.10\times10^{-11}T_e^{-1.48}\exp(-0.68/T_e)$ | 0.14 | 19,20 |
| 10 | $e+Cl_2 \rightarrow e+Cl_2(v3)$ | $2.39\times10^{-11}T_e^{-1.49}\exp(-0.64/T_e)$ | 0.21 | 19,20 |
| 11 | $e+Cl_2(v1) \rightarrow e+Cl_2(v2)$ | $1.04\times10^{-9}T_e^{-1.48}\exp(-0.73/T_e)$ | 0.07 | 19,20 |
| 12 | $e+Cl_2(v1) \rightarrow e+Cl_2(v3)$ | $2.98\times10^{-10}T_e^{-1.48}\exp(-0.67/T_e)$ | 0.14 | 19,20 |
| 13 | $e+Cl_2(v2) \rightarrow e+Cl_2(v3)$ | $1.04\times10^{-9}T_e^{-1.48}\exp(-0.73/T_e)$ | 0.07 | 19,20 |
| 14 | $e+Cl_2 \rightarrow 2Cl+e$ | $6.67\times10^{-8}T_e^{-0.1}\exp(-8.67/T_e)$ | 4 | 19,20 |
| 15 | $e+Cl_2 \rightarrow Cl_2^++2e$ | $4.87\times10^{-8}T_e^{0.5}\exp(-12.17/T_e)$ | 11.5 | 19,20 |
| 16 | $e+Cl_2 \rightarrow Cl+Cl^++2e$ | $1.79\times10^{-7}\exp(-24.88/T_e)$ | 14.25 | 19,20 |
| 17 | $e + Cl_2 \rightarrow 2Cl^+ + 3e$ | $1.46\times10^{-10} T_e^{2.16} \exp(-21.42/T_e)$ | 28.5 | 19,20 |
| 18 | $e + Cl_2 \rightarrow Cl^+ + Cl^- + e$ | $3.45\times10^{-10} T_e^{0.13} \exp(-19.7/T_e)$ | 14.25 | 19,20 |
| 19 | $e + Cl_2 \rightarrow Cl + Cl^-$ | $\begin{pmatrix} 22.5 T_e^{-0.46} e^{-2.82/T_e} \\ -12.1 e^{-0.99/T_e} + 6.54 \end{pmatrix} \times10^{-10}$ | | 19,20 |
| 20 | $e + Cl_2(v1) \rightarrow Cl + Cl^-$ | $\begin{pmatrix} 9.29 T_e^{-0.47} e^{-2.83/T_e} \\ -4.96 e^{-0.99/T_e} + 2.7 \end{pmatrix} \times10^{-9}$ | | 19,20 |
| 21 | $e + Cl_2(v2) \rightarrow Cl + Cl^-$ | $\begin{pmatrix} 20.1 T_e^{-0.47} e^{-2.83/T_e} \\ -10.8 e^{-0.97/T_e} + 5.92 \end{pmatrix} \times10^{-9}$ | | 19,20 |
| 22 | $e + Cl_2(v3) \rightarrow Cl + Cl^-$ | $\begin{pmatrix} 30.5 T_e^{-0.46} e^{-2.82/T_e} \\ -16.3 e^{-0.99/T_e} + 8.81 \end{pmatrix} \times10^{-9}$ | | 19,20 |
| 23 | $e + Cl \rightarrow Cl^+ + 2e$ | $2.48\times10^{-8} T_e^{0.62} \exp(-12.76/T_e)$ | 14.25 | 19,20 |
| 24 | $e + Cl^- \rightarrow Cl + 2e$ | $2.33\times10^{-15} T_e^{1.45} \exp(-2.48/T_e)$ | 2.36 | 19,20 |
| 25 | $e + Cl^- \rightarrow Cl^+ + 3e$ | $3.38\times10^{-9} T_e^{0.75} \exp(-25.28/T_e)$ | 16.61 | 19,20 |
| 26 | $e + Cl_2^+ \rightarrow 2Cl$ | $9.0\times10^{-8} T_e^{-0.5}$ | -11.5 | 19,20 |
| 27 | $Ar^m + Ar^m \rightarrow e + Ar + Ar^+$ | $6.2\times10^{-10}$ | 0 | 19,20 |
| 28 | $Ar^m + Ar \rightarrow Ar + Ar$ | $3.0\times10^{-15}$ | 0 | 19,20 |
| 29 | $Cl_2^+ + Cl^- \rightarrow 3Cl$ | $5.0\times10^{-8} (300/T_g)^{0.5}$ | 0 | 19,20 |
| 30 | $Cl_2^+ + Cl^- \rightarrow Cl + Cl_2$ | $5.0\times10^{-8} (300/T_g)^{0.5}$ | 0 | 19,20 |
| 31 | $Cl^+ + Cl^- \rightarrow 2Cl$ | $5.0\times10^{-8} (300/T_g)^{0.5}$ | 0 | 19,20 |
| 32 | $Cl^+ + Cl_2 \rightarrow Cl + Cl_2^+$ | $5.4\times10^{-10}$ | 0 | 19,20 |
| 33 | $2Cl + Cl_2 \rightarrow Cl_2 + Cl_2$ | $3.5\times10^{-39} \exp(810/T_g)$ | 0 | 19,20 |
| 34 | $2Cl + Cl \rightarrow Cl_2 + Cl$ | $8.75\times10^{-40} \exp(810/T_g)$ | 0 | 19,20 |
| 35 | $Cl_2 + Ar^+ \rightarrow Cl_2^+ + Ar$ | $1.9\times10^{-10}$ | 0 | 19,20 |
| 36 | $Cl_2 + Ar^+ \rightarrow Cl + Cl^+ + Ar$ | $5.7\times10^{-10}$ | 0 | 19,20 |
| 37 | $Cl^- + Ar^+ \rightarrow Cl + Ar$ | $5.0\times10^{-8} (300/T_g)^{0.5}$ | 0 | 19,20 |
| 38 | $Cl_2 + Ar^m \rightarrow Cl_2^+ + Ar + e$ | $7.1\times10^{-10}$ | 0 | 19,20 |
| 39 | $Cl + Ar^+ \rightarrow Cl^+ + Ar$ | $2.0\times10^{-10}$ | 0 | 19,20 |
| 40 | $2Cl + Ar \rightarrow Cl_2 + Ar$ | $8.75\times10^{-40}\exp\left(-810/T_g\right)$ | 0 | 19,20 |
| 41 | $2Cl + Ar^m \rightarrow Cl_2 + Ar$ | $8.75\times10^{-40}\exp\left(-810/T_g\right)$ | 0 | 19,20 |

[a]The unit of the rate coefficient is $cm^3s^{-1}$.

[b] v1,v2,v3 represent different vibrational excited states.

Table 2. Surface reaction set considered in the model

| No. | Surface reaction | Sticking coefficient | Ref. |
|---|---|---|---|
| 1 | $Cl_2(v1) + wall \rightarrow Cl_2$ | 1 | 19,20 |
| 2 | $Cl_2(v2) + wall \rightarrow Cl_2$ | 1 | 19,20 |
| 3 | $Cl_2(v3) + wall \rightarrow Cl_2$ | 1 | 19,20 |
| 4 | $Cl_2^+ + wall \rightarrow Cl_2$ | 1 | 19,20 |
| 5 | $Cl^+ + wall \rightarrow Cl$ | 1 | 19,20 |
| 6 | $Cl + wall \rightarrow 1/2\, Cl_2$ | 0.04 | 21 |
| 7 | $Ar^+ + wall \rightarrow Ar$ | 1 | 19,20 |
| 8 | $Ar^m + wall \rightarrow Ar$ | 1 | 19,20 |

## III. Results and discussion

~~~~~~~~~~~~~~~~~~~~~~~~~~~~~~~~~~~~~~~~~~~~~~~~~~~~~~~~~~~~~~~~~~~~~~~~

Catalogue of section III:

(3.1) Increasing $Cl_2$ content at low pressures, *e.g.*, 10mTorr, from 5% $Cl_2$ content
- (a) Species-stratification, quasi-delta anion and self-coagulation
- (b) Conventional space-stratification analogous to Ar/$SF_6$ discharge
- (c) From delta/comet profile to parabola and the core expansion: $Cl_2$ gas ratio effect

(3.2) Increasing the pressure at high $Cl_2$ contents, *e.g.*, 90%, from 10mTorr
- (a) Evolution of ions density profile from parabola to flat-top, besides for self-coagulation-to-coil (SCC) scheme
- (b) Electron coagulates at ambi-polar diffusion and ions self-coagulation
- (c) Ambi-polar diffusion of triple-species (*i.e.*, *gentle* ambi-polar self-coagulation)
- (d) Discussion on plasma non- electric neutrality and its collective interaction

(3.3) Decreasing the $Cl_2$ content at high pressures, *e.g.*, 90mTorr, from 90% $Cl_2$ content
- (a) Collapse of self-coagulation-to-coil
- (b) Hollow anion density and grouping effect
~~~~~~~~~~~~~~~~~~~~~~~~~~~~~~~~~~~~~~~~~~~~~~~~~~~~~~~~~~~~~~~~~~~~~~~~

(c) Reactive feedstock gas depletion and shortage
(d) Electron coagulates at ambi-polar diffusion, without any self-coagulation

(3.4) Decreasing the pressure at low $Cl_2$ contents, *e.g.*, 5%, from 90mTorr
(a) From hollow to $\delta$ type and disappearance of grouping
(b) Refreshment of reactive feedstock gas and de-coagulation of electron

~~~~~~~~~~~~~~~~~~~~~~~~~~~~~~~~~~~~~~~~~~~~~~~~~~~~~~~~~~~~~~~~~~~~~~~~~~~~

(3.1) Increasing $Cl_2$ content at low pressures, *e.g.*, 10mTorr, from 5% $Cl_2$ content
(a) Species-stratification, quasi-delta anion and self-coagulation

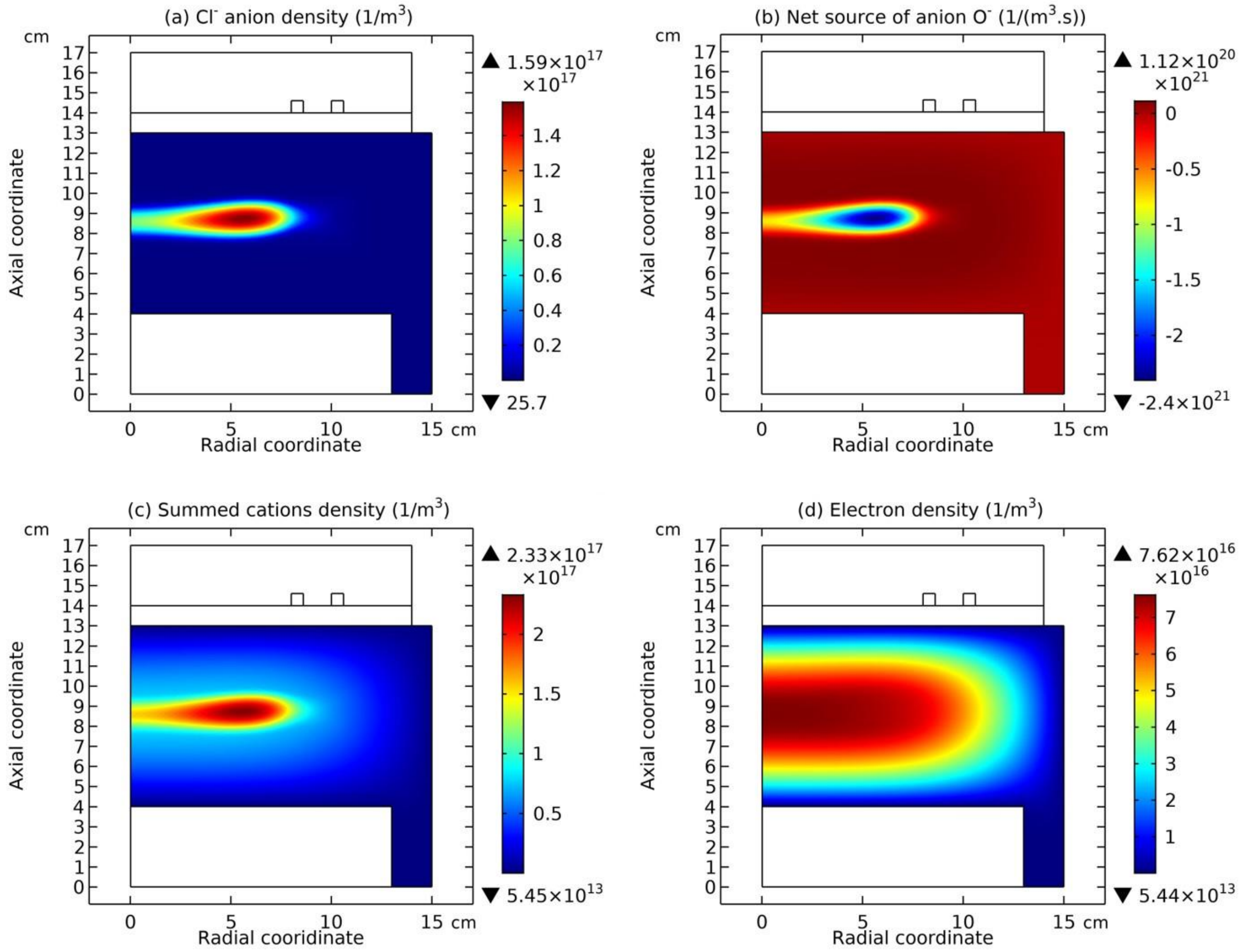

Figure 1 $Cl^-$ anion density (a), the net source of $Cl^-$ (b), summed cations density (c) and electron density (d) in the $Ar/Cl_2$ inductively coupled plasma, given by the fluid simulation, at the discharge conditions of 300W, 10mTorr and 5% of $Cl_2$ gas content.

As the early analytic work predicted, there are two types of discharge stratification in the electronegative plasmas. Here in this section, the known species-stratification is illustrated, which occurs at the case of low electronegativity. It indicates that the densities of different charged species in the electronegative core is stratified [8]. This is different with the so-called space-stratification that divides the whole bulk discharge region into the electronegative core and electropositive edge [9]. The fluid simulation shown in Fig. 1 gives the $Cl^-$ anion density and its net source, as well as the summed cations density and electron density, at only 5% $Cl_2$ content in the feedstock gas mixture. As seen, the $Cl^-$ density behaves like a comet, also called as quasi-delta profile. This novel anion density distribution has been reported in our previous paper about $Ar/O_2$ inductively coupled plasma [17], which is a low electronegativity case as well. It was interpreted by the self-coagulation behavior that is controlled by the quasi- Helmholtz equation. Self-coagulation we reported is one type of self-organization behaviors. It naturally happens once the necessary conditions are
~~~~~~~~~~~~~~~~~~~~~~~~~~~~~~~~~~~~~~~~~~~~~~~~~~~~~~~~~~~~~~~~~~~~~~~~~~~~

satisfied, *i.e.*, negative chemical source and transport component of free diffusion. Superficially, the self-consistent fluid simulation predicts a delta profile, not in accord to the analytic solution that gives the smooth stratified species density profiles. Nevertheless, when zooming into the delta inside, it is found that species-stratification is hidden by the delta structure in Fig. 2. As seen, after the operations of expansion, normalization and plotting format revision, the fluid simulation perfectly repeats the analytic work, in the scope of species-stratification. In the simulation, why the delta structure is superposed? Caused by the anion dynamics (described above). The anion is assumed to satisfy the Boltzmann relation in the analytics and the self-organization cannot be considered thereby. Although the simulation is more complete, the importance of analytic work predictions is emphasized, since it is difficult to understand the simulations without the guide of analytics.

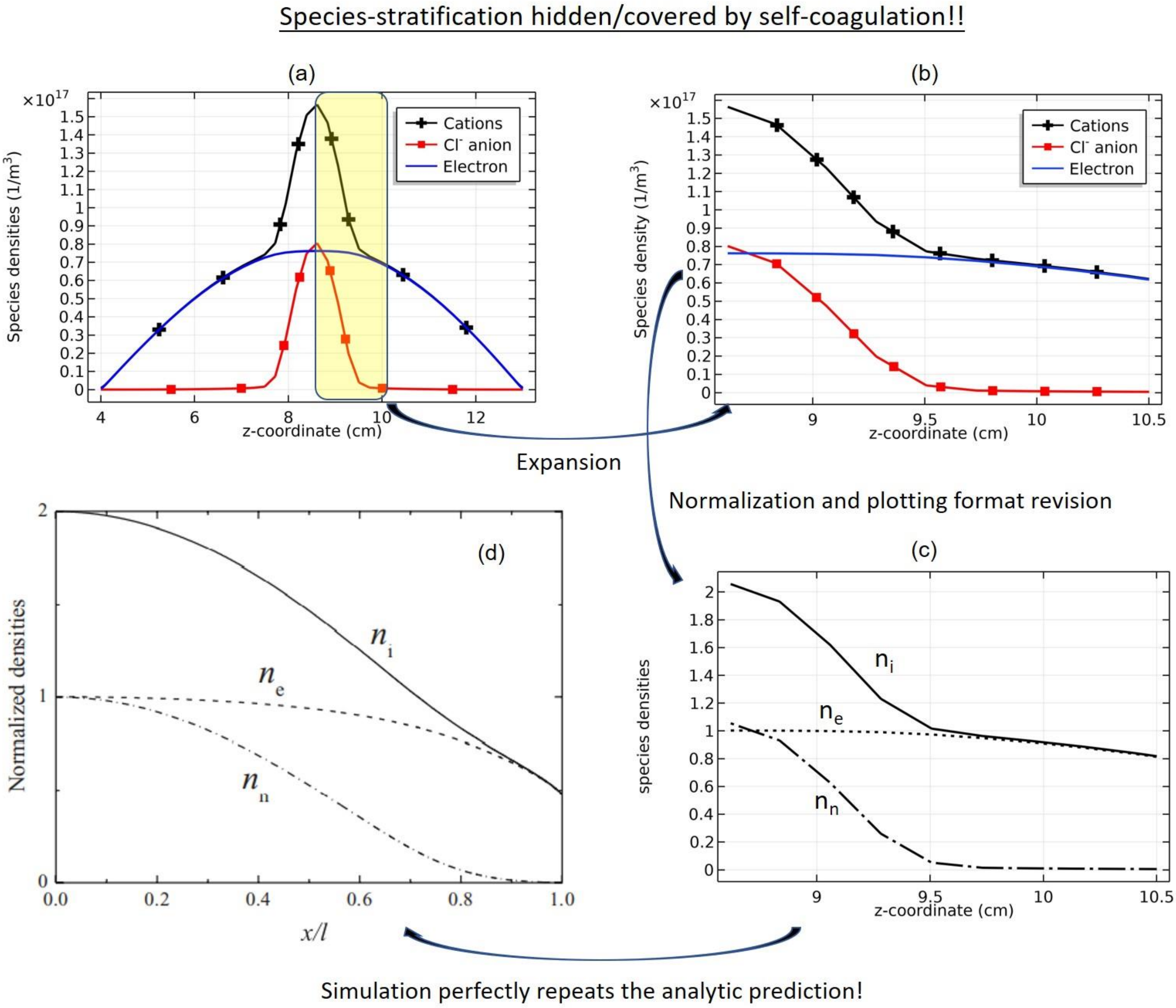

Figure 2 Simulated axial profiles of different species densities of $Ar/Cl_2$ inductive discharge along the reactor axis (a), the expanded part of these axial profiles inside the half delta structure (b), the expanded profiles plotted at different format and normalized (c), and the species density profiles of analytic solution that ignores the anion dynamics [8] (d). The simulation is given by the fluid model at the discharge conditions of 300W, 10mTorr and 5% $Cl_2$ content. This plot shows that the species-stratification is hidden by the self-coagulation behavior.

### (b) Conventional space-stratification analogous to Ar/$SF_6$ discharge

At high enough $Cl_2$ content ratio, 90%, the discharge structure of Ar/$Cl_2$ inductive plasma evolves to the conventional space-stratification, *i.e.*, the combination of electronegative core and electropositive edge, as shown in Fig. 3. Fig. 4 plots the axial and radial distributions of species densities, respectively. As seen, this is very similar to the Ar/$SF_6$ inductive discharge case, where the parabola trait of core and the self-coagulation portion along the radial profile are displayed. The difference is that the Ar/$Cl_2$ plasma electronegativity is weaker than the Ar/$SF_6$ plasma. Hence, the reactive gas content of Ar/$Cl_2$ plasma (90%) here is higher than the Ar/$SF_6$ case, *i.e.*, 10%. Even at high reactive gas ratio, the electronegativity is slightly lower than the Ar/$SF_6$ plasma. Accordingly, the electropositive halo of Ar/$Cl_2$ plasma is wider and the core electron density is higher, upon comparing with Ar/$SF_6$ plasma, in accord to the analytic prediction [7].

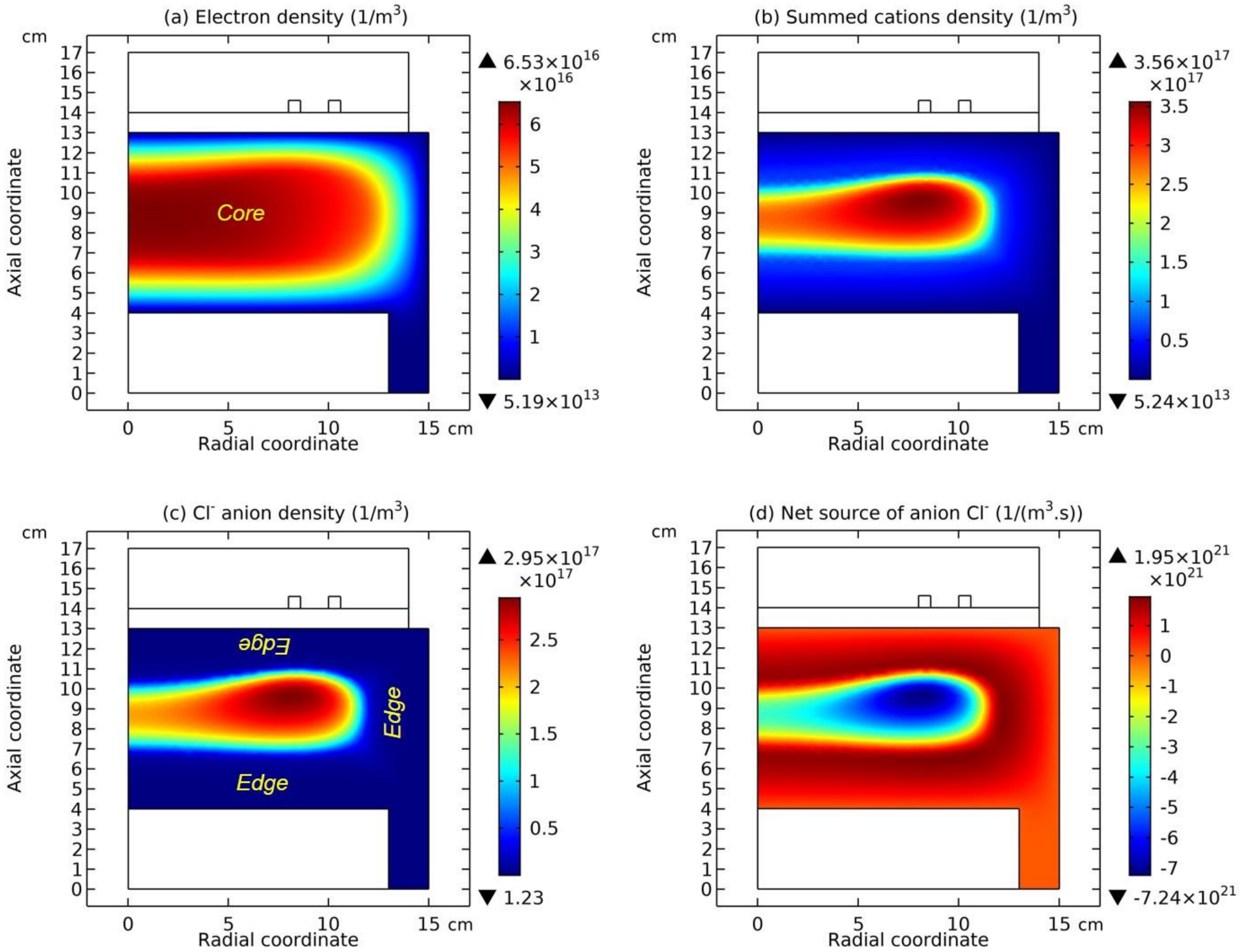

Figure 3 Electron density (a), summed cations density (b), $Cl^-$ anion density (c), and its net source (d) of Ar/$Cl_2$ inductively coupled plasma given by the fluid simulation at the discharge conditions of 300W, 10mTorr and 90% $Cl_2$ content.

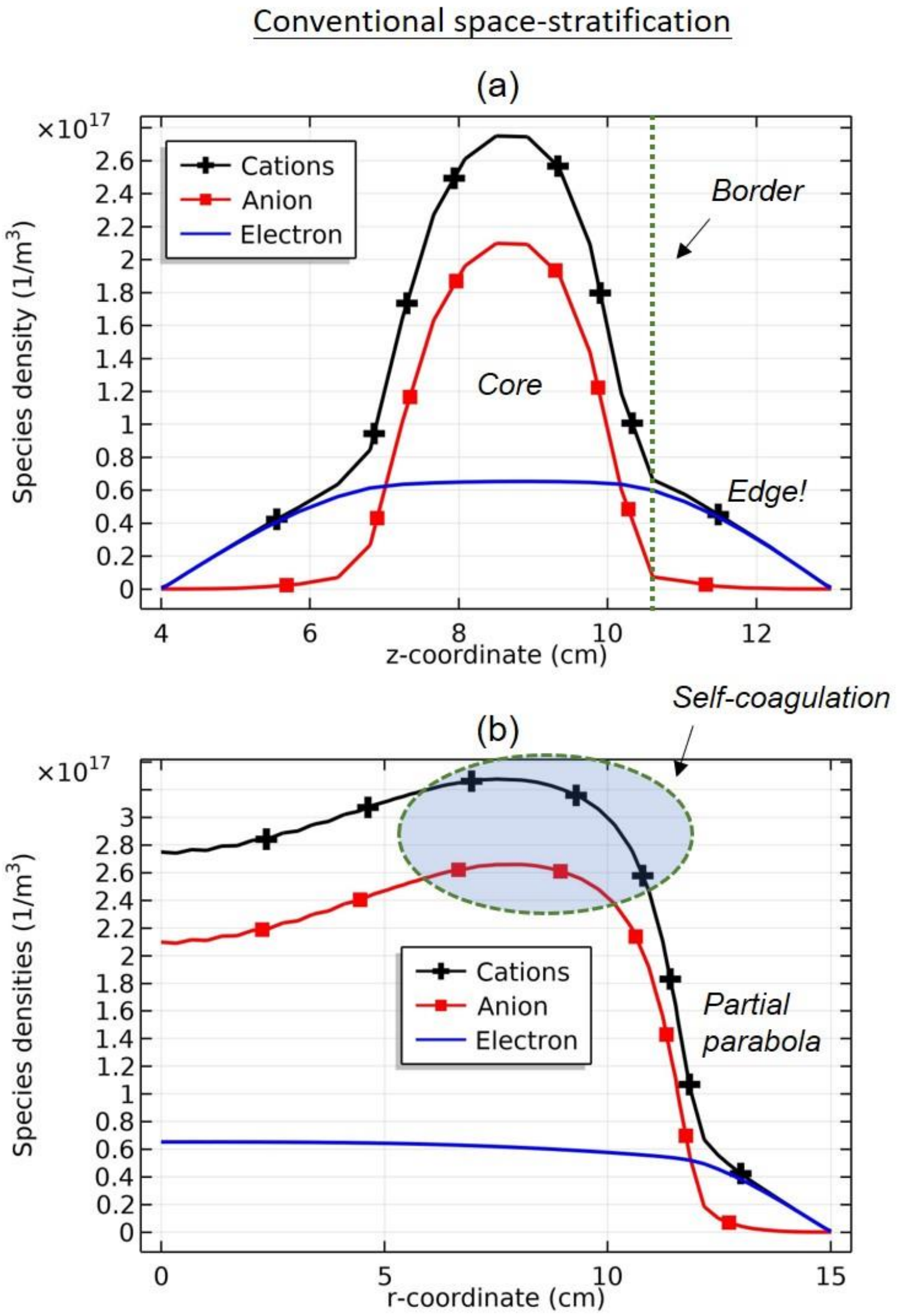

Figure 4 Axial (a) and radial (b) profiles of species densities of Ar/$Cl_2$ inductive discharge given by the fluid simulation at the discharge conditions of 300W, 10mTorr and 90% $Cl_2$ content. This figure displays the conventional space-stratification structure and the embodied self-coagulation structure.

### (c) From delta/comet profile to parabola and the core expansion: $Cl_2$ gas content effect

The simulations show that delta plasma profile of Ar/$Cl_2$ inductive discharge at 10mTorr evolves into the parabola profile upon increasing the $Cl_2$ gas ratio, as shown in Fig. 5. At low $Cl_2$ content, the delta profile is a special *space-stratification structure* at the extreme condition, very low electronegativity, that the electronegative core is substantially shrunk. This is logic and determined by the origin of space-stratification [16]. At the two cases, the self-coagulation is accompanied. The difference is that at low $Cl_2$ content, electronegative core is *slim* and self-coagulation is predominant; and so, anion displays the comet profile. While at high $Cl_2$ content, core is *fat* and self-coagulation is embodied in the wide parabola profile.

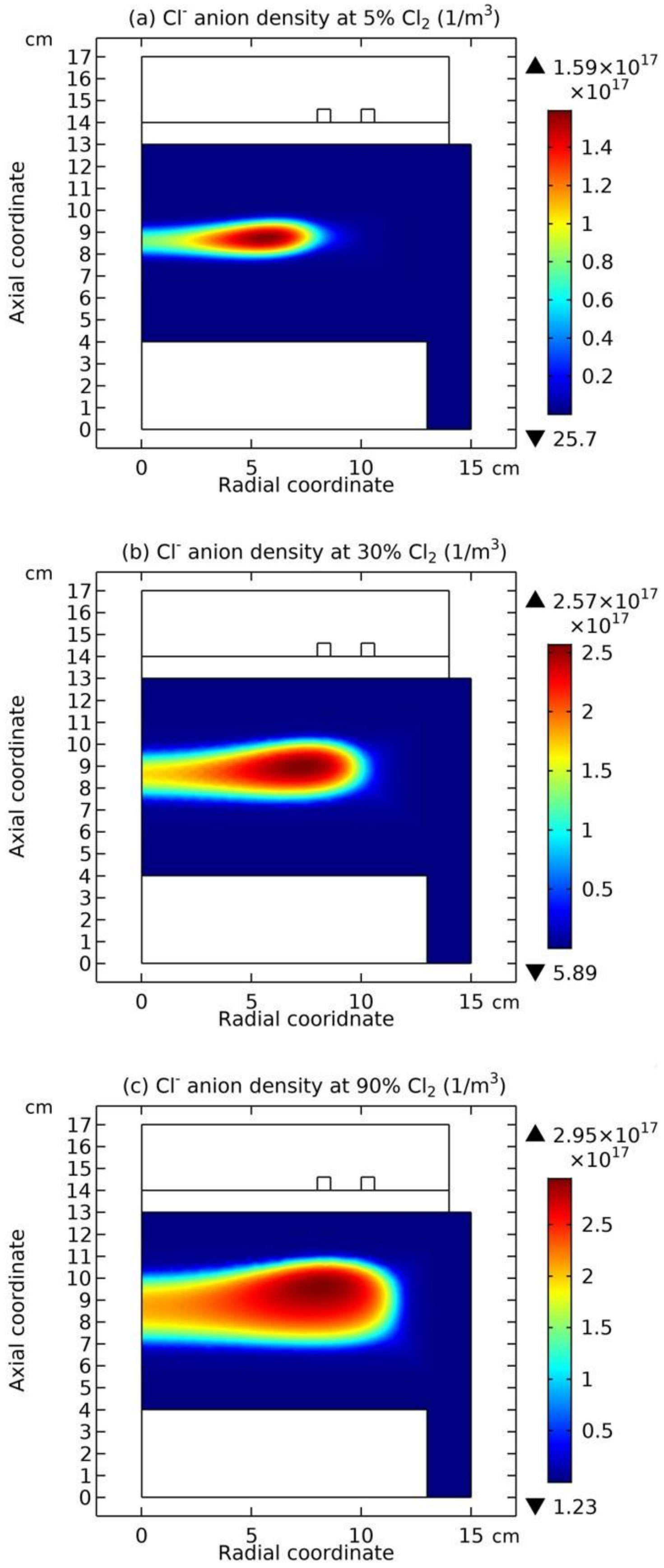

Figure 5 $Cl^-$ anion density at (a) 5%, (b) 30% and (c) 90% $Cl_2$ contents, given by the fluid simulation. The other discharge conditions are 300W and 10mTorr. This plot illustrates the evolution of delta profile toward the parabola profile, upon increasing the reactive gas ratio.

(3.2) Increasing the pressure at high $Cl_2$ contents, *e.g.*, 90%, from 10mTorr

(a) Evolution of ions density profile from parabola to flat-top, besides for self-coagulation-to-coil (SCC) scheme

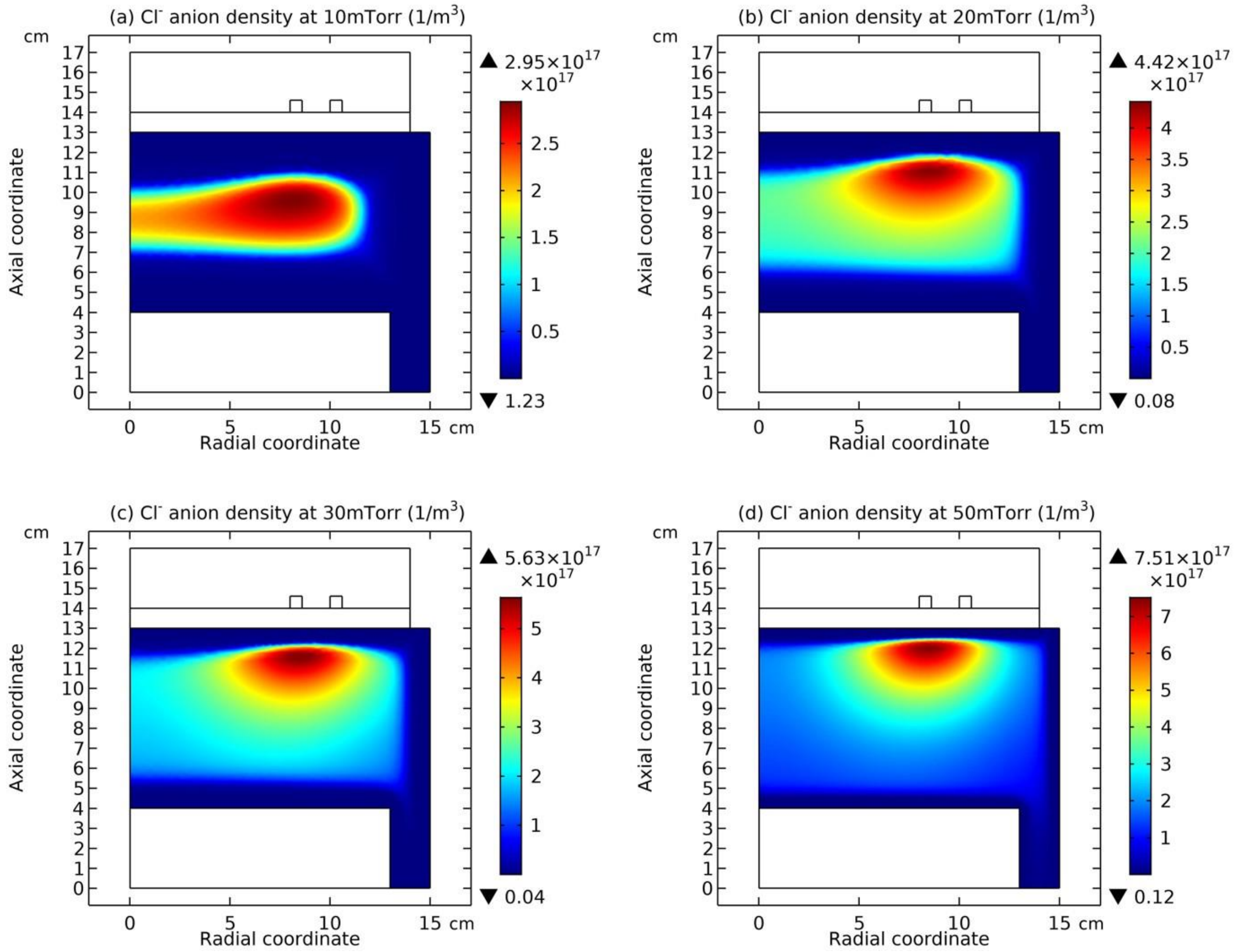

Figure 6 $Cl^-$ anion density at (a) 10mTorr, (b) 20mTorr, (c) 30mTorr and (d) 50mTorr, at 90% $Cl_2$ content and 300W, simulated by the fluid model.

In Fig. 6, the evolution of $Cl^-$ anion density with pressure is presented at fixed $Cl_2$ content (90%) and 300W. Besides for the self-coagulation-to-coil (SCC) behavior (reported in Ref. [16]), it experiences the parabola, ellipse and flat-top models (the same as in the $Ar/SF_6$ plasma [16]), which is already well-known and determined by their respective analytic theory [16]. The cations density displays the same trend against pressure, given by the ambi-polar diffusion potential which is not collapsed. See the Secs. (b, c) for more details. This is different with the $Ar/SF_6$ plasma where the *advective* ambi-polar self-coagulation scheme, *i.e.*, blue negative internal sheath, helps the cations coagulate [16].

(b) Electron coagulates at ambi-polar diffusion and ions self-coagulation

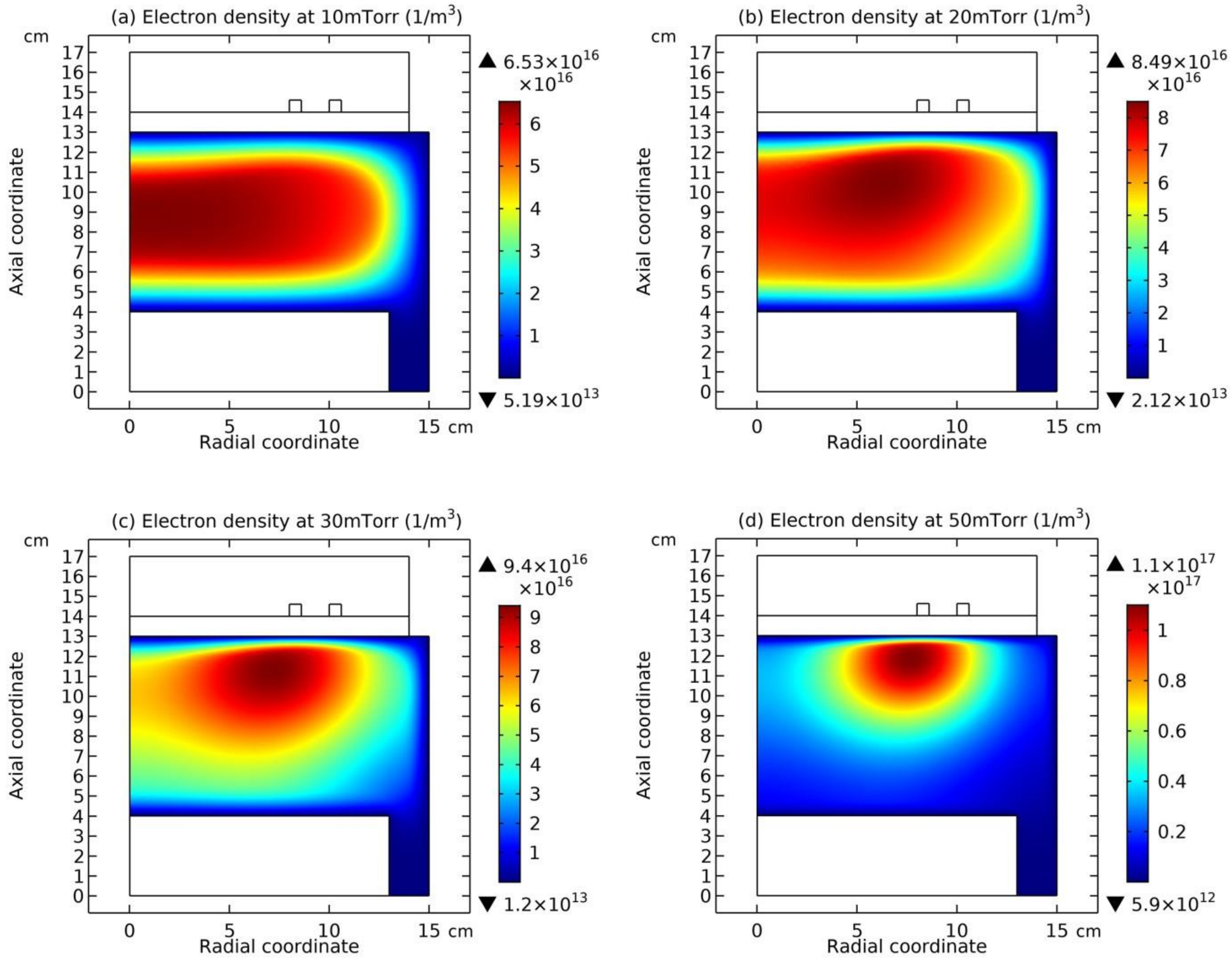

Figure 7 Electron density profiles of Ar/$Cl_2$ inductive plasma at different pressures, (a) 10mTorr, (b) 20mTorr, (c) 30mTorr, and (d) 50mTorr, given by fluid simulation at 300W and 90% $Cl_2$ content.

In Fig. 7, the electron density coagulates toward the coil at increasing the pressure because of the ions self-coagulation, at 90% $Cl_2$ content and 300W power, very similar to the Ar/$SF_6$ plasma [16]. The difference is that in the Ar/$Cl_2$ plasma, the ambi-polar diffusion potential of high pressure, *e.g.*, 50mTorr, is not collapsed, as shown in Fig. 8(b), due to the low electronegativity exhibited in Fig. 9. Since free-diffusion condition is not satisfied, the peripherical self-coagulation of electrons is not happened and hence electron is Boltzmann balanced (also quite different with Ar/$SF_6$ plasma) in Fig. 8(c). The similar trend in 90mTorr Ar/$Cl_2$ plasma of 90% $Cl_2$ and 300W is presented in Fig. 8(d-f). It is seen that the anion's self-coagulation leads to the electron's *tight coagulation* to the coil, while the electron's self-coagulation deviates the Boltzmann relation. At the high pressure of 90mTorr, when decreasing the $Cl_2$ content to 5%, this *tight-coagulation-to-coil* of electron (a high electronegativity behavior), *i.e.*, with its density peak almost *attached* to the dielectric window, is replaced by a departing-coil-type coagulation, where anion self-coagulation is destroyed at the shortage of reactive feedstock. See the Fig. 22 of Sec. (3.3) for reference.

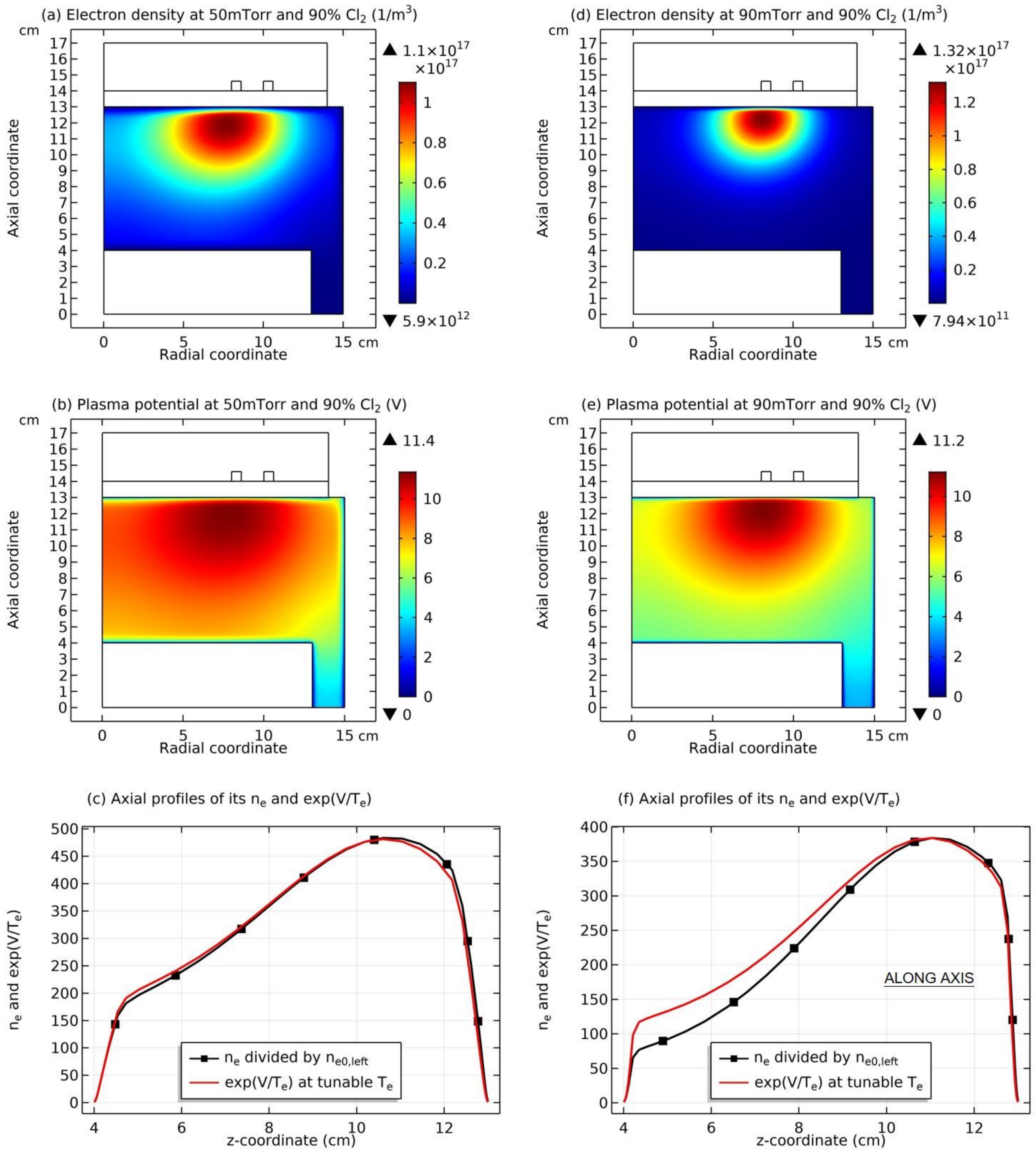

Figure 8 Electron density (a), plasma potential (b) and its Boltzmann relation (c) of 50mTorr inductive $Ar/Cl_2$ plasma, and these three quantities (d-f) of 90mTorr inductive $Ar/Cl_2$ plasma, given by fluid simulation at 300W and 90% $Cl_2$ content.

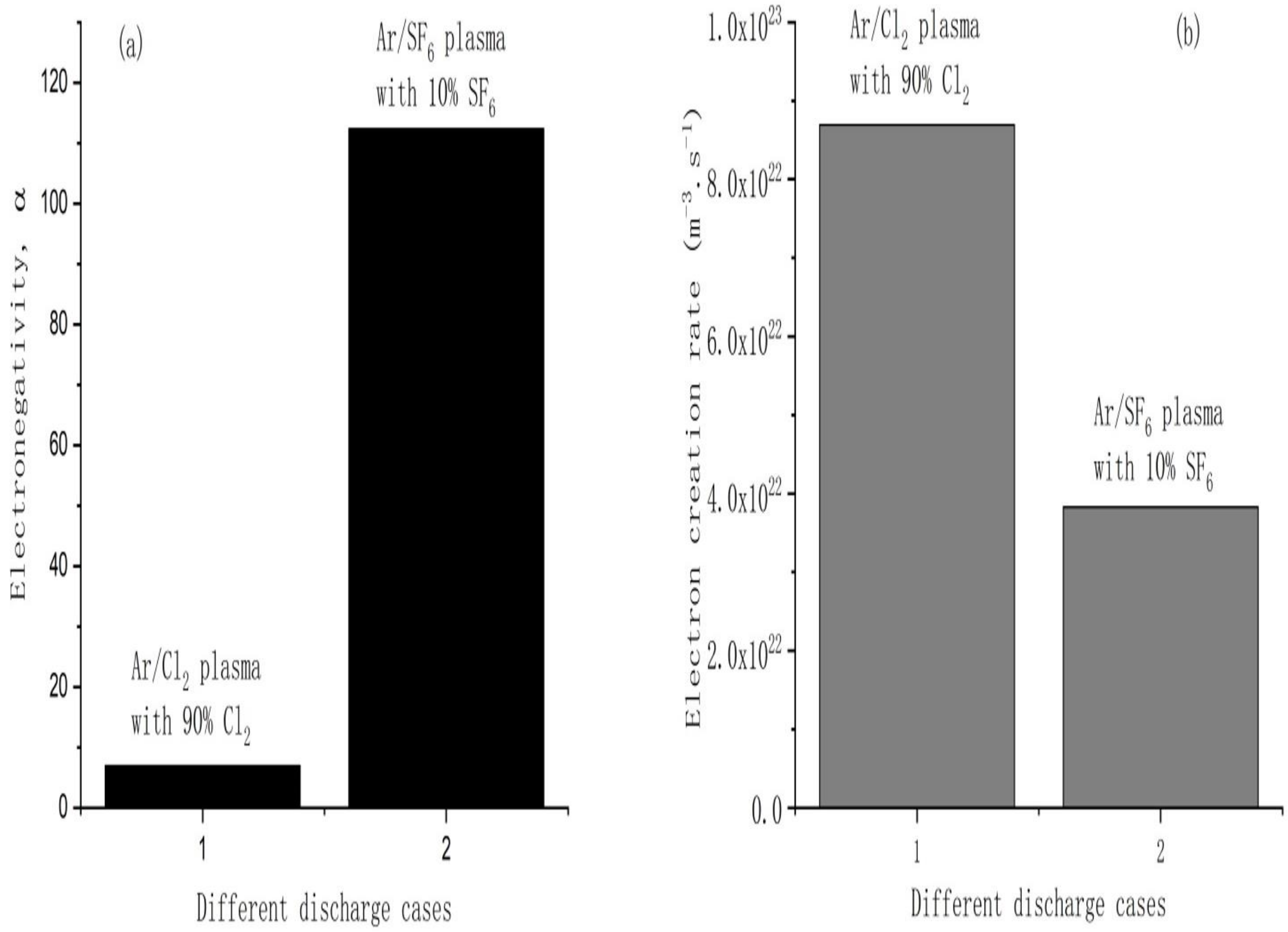

Figure 9 Comparison of electronegativities (a) and electron creation sources (b) between the Ar/$Cl_2$ and Ar/$SF_6$ plasmas, at 50mTorr. The electronegativity is defined as $\alpha = \frac{n_-}{n_e}$, where $n_-, n_e$ are the total anions density and electron density, respectively. The electronegativity is low at Ar/$Cl_2$ plasma, *i.e.*, 6.83 given at the condition of $n_- = 7.51\times10^{17}\ m^{-3}, n_e = 1.1\times10^{17}\ m^{-3}$. The electronegativity is high in Ar/$SF_6$ plasma, *i.e.*, 112.3 given at the condition of $n_- = 2.46\times10^{18}\ m^{-3}, n_e = 2.19\times10^{16}\ m^{-3}$. Electron creation rate in Ar/$Cl_2$ plasma is higher than the Ar/$SF_6$, implying its low electronegativity.

### (c) Ambi-polar diffusion of triple-species (*i.e.*, *gentle* ambi-polar self-coagulation)

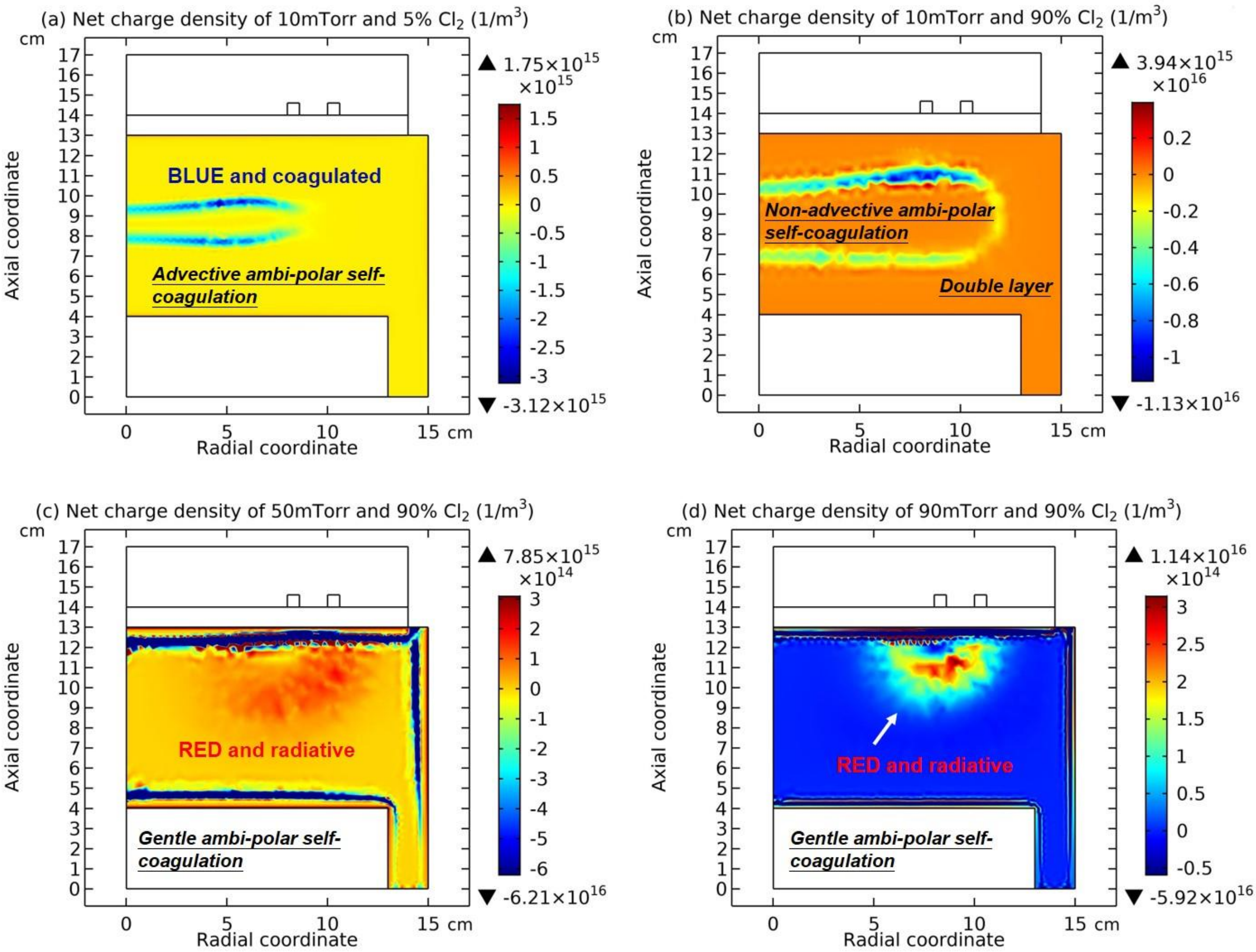

Figure 10 Net charge density of Ar/$Cl_2$ inductive plasma at (a) 10mTorr and 5% $Cl_2$ content, (b) 10mTorr and 90% $Cl_2$ content, (c) 50mTorr and 90% $Cl_2$ content, and (d) 90mTorr and 90% $Cl_2$ content, given by fluid simulation at 300W power.

In this section, the ambi-polar self-coagulation of Ar/$Cl_2$ plasma is discussed. In Fig. 10, the net charge densities of different discharge conditions of this plasma are presented. In Fig. 10(a), at 10mTorr and 5% $Cl_2$ content, where delta type anion is given, the blue negative sheath is seen, which represents the advective ambi-polar self-coagulation, as illustrated in Ref. [16] when the Ar/$O_2$ inductive plasma is involved. The non-advective (*i.e.*, spontaneous) ambi-polar self-coagulation is observed at 10mTorr and 90% $Cl_2$ content, like the case of Ar/$SF_6$ plasma at 10mTorr and 10% $SF_6$ content [16]. Of interest is that at 50mTorr and 90mTorr of 90% $Cl_2$ content, it is so strange to find that the blue negative sheath is not appeared, different with the Ar/$SF_6$ plasma at similar conditions (compare to the Fig. 51(c,d) of Ref .[16]). This is because that the ambi-polar diffusion potential of Ar/$Cl_2$ plasma is not collapsed and it helps the cations gently following the anion's self-coagulation, not forming sheath. This is analogous to the electropositive plasma where ion gently follows the electron's self-diffusion at ambi-polar diffusion potential, which is approximated electrically neutral. In a word, the process of ambi-polar diffusion is known to eliminate the non-neutrality of plasma, suitable for the escaping electrons (via free-diffusion) and anions (via chemical coagulation) relative to the cation coordinate, as shown in Fig. 11. So, this is called *gentle* ambi-polar self-coagulation. The red and radiative sheaths of 50mTorr and 90mTorr cases are originated from the mass-point expelling effect, as has been illustrated in Ref. [16]. This non-neutrality is not that strong as in Ar/$SF_6$ plasma, as compared with Fig. 41(d) of Ref. [16], since the density orders of different cations are not close to each other, as shown in Fig. 12.

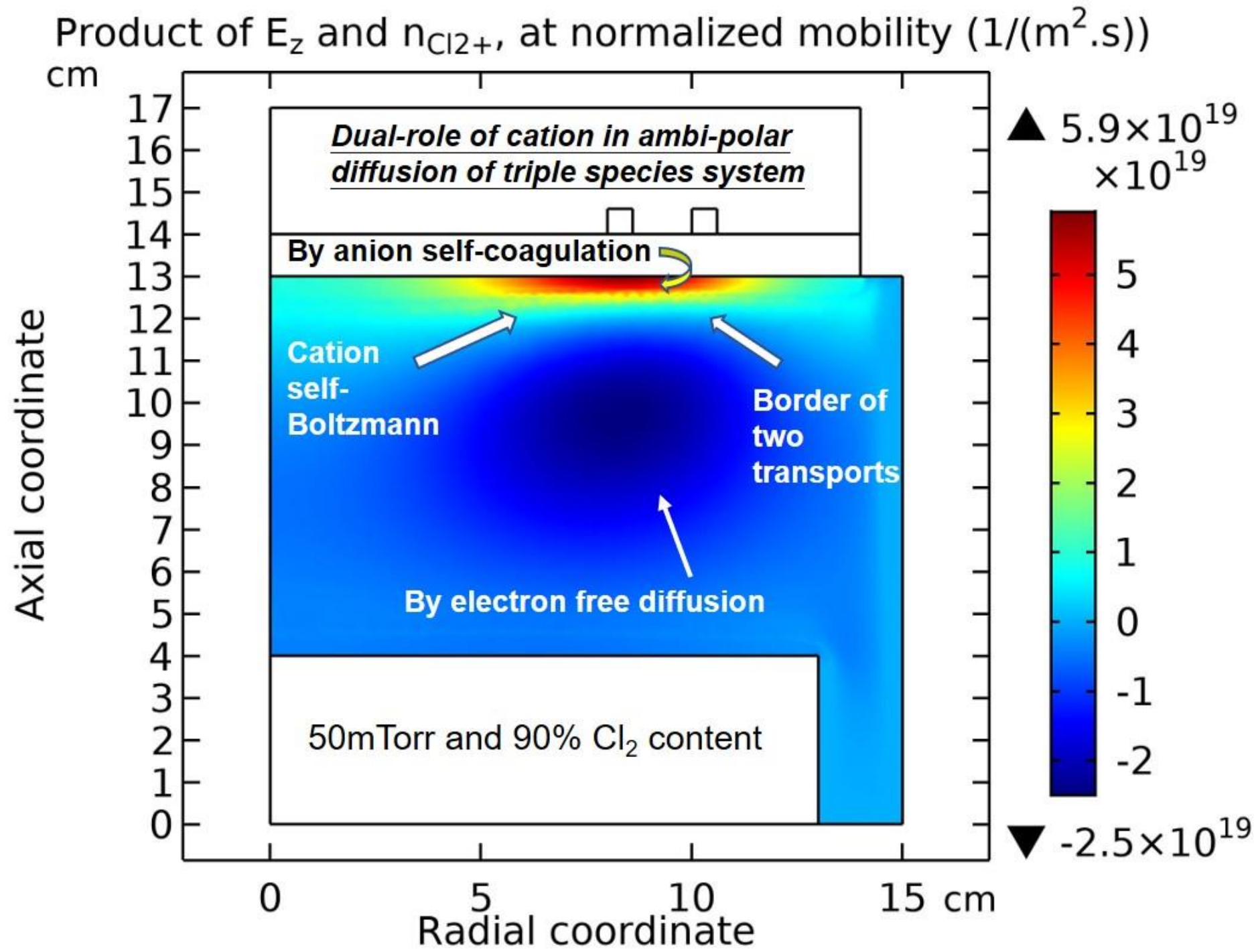

Figure 11 Dual-role of cation in the ambi-polar diffusion of triple species system, in the quantity of axial electric field and $Cl_2^+$ density product (at normalized mobility) of 50mTorr and 90% $Cl_2$ content inductive Ar/$Cl_2$ plasma, given by fluid model simulation at 300W power. The 90mTorr Ar/$Cl_2$ plasma exhibits the same trend. The product of electric field and cation density represents the drift of it in the ambi-polar diffusion. As the analytic work predicted [5], the drift of ion is the essence of its transport in the process of ambi-polar diffusion of two species system (electron and ion). Other cations, $Ar^+$ and $Cl^+$, exhibit the same trend as $Cl_2^+$. The dual-role of cation in the process is originated from electron free diffusion and anion self-coagulation. Free diffusion directs to substrate and self-coagulation directs to dielectric window. So, the drift of cations consists of two parts, thin red part and wide blue part (refer to the cylindrical coordinate system). At the interface (or border) of the two transports, the cation drift is almost zero at the counteract of ambi-polar diffusion and self-coagulation, indicating its self-Boltzmann property.

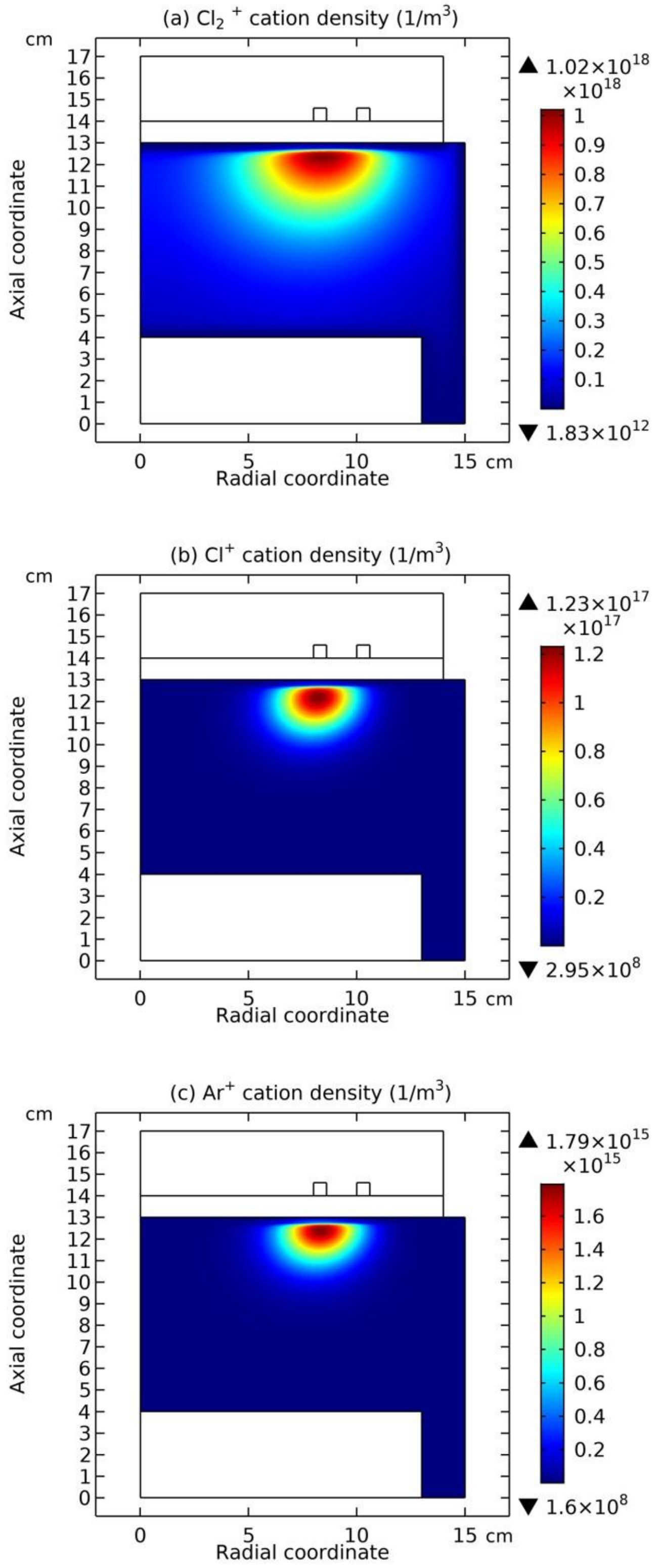

Figure 12 Densities of three main cations, $Cl_2^+$, $Cl^+$ and $Ar^+$ in the $Ar/Cl_2$ inductive plasma of 90mTorr and 90% $Cl_2$ content, given by fluid simulation at 300W.

(d) Discussion on plasma non- electric neutrality and its collective interaction

Table 3 Summary of plasma non-neutrality (left column) and quasi-neutrality (right column) discovered in low-temperature plasma sources at present.

| Plasma obvious/strong non-neutrality | | Plasma quasi-neutrality: *very weak and negligible charge* | |
|---|---|---|---|
| **Collective role** | **Against role** | **Collective role** | |
| Red sheath | Double layer | In electropositive plasmas | In electronegative plasmas |
| Blue sheath | Mass point potential | Ion-electron ambi-polar diffusion | ✧ Cation, anion and electron ambi-polar diffusion<br>✧ Gentle or/and spontaneous ambi-polar self-coagulation<br>✧ Conventional cation-electron ambi-polar diffusion |

In this section, the non-neutralities discovered presently in the low temperature plasma sources are summarized and discussed. Based on our understanding, they are divided into two parts, strong non-neutrality and quasi-neutrality, as illustrated in Tab. 3. Furthermore, the strong non-neutrality itself is divided into two parts, which are given by collective and against roles, respectively. The plasma quasi-neutrality is focused on the very weak and almost negligible charge accumulation. For now, it is given by the collective role. It includes two parts as well, which are happened respectively in electropositive and electronegative plasmas. In electropositive plasma, the ion-electron dual-species ambi-polar diffusion is considered, and in electronegative plasma the triple-species (cation, anion and electron) ambi-polar diffusion is suggested, which actually contains the cation-anion self-coagulation and cation-electron diffusion (both are ambi-polar).

Collective role means that the various plasma species interact with each other for finishing certain mission, *i.e.*, a collaboration behavior. Many instances belong to such type, *e.g.*, chamber border sheath (positive, always plotted red in the colormap). Besides, the advective ambi-polar self-coagulation early reported in Ref. [16] generates electronegative sheath in the plasma bulk, which is plotted as blue in the colormap. On the opposite, against role implies that certain processes of complex plasma fight against each other, and finally give rise to special consequence. For instance, the *mass point potential*, given by the expelling effect of transformed charge elements with same polarity (originated from self-coagulation), and *the double layer*, produced by the pushing and accumulating anion of ambi-polar potential of cation and electron. As analyzed, these interactions are conflicted and therefore we propose the definition of against role, relative to the collective role (well-known property), for better understanding plasma. More details of these plasma non-neutralities are given in Fig. 13.

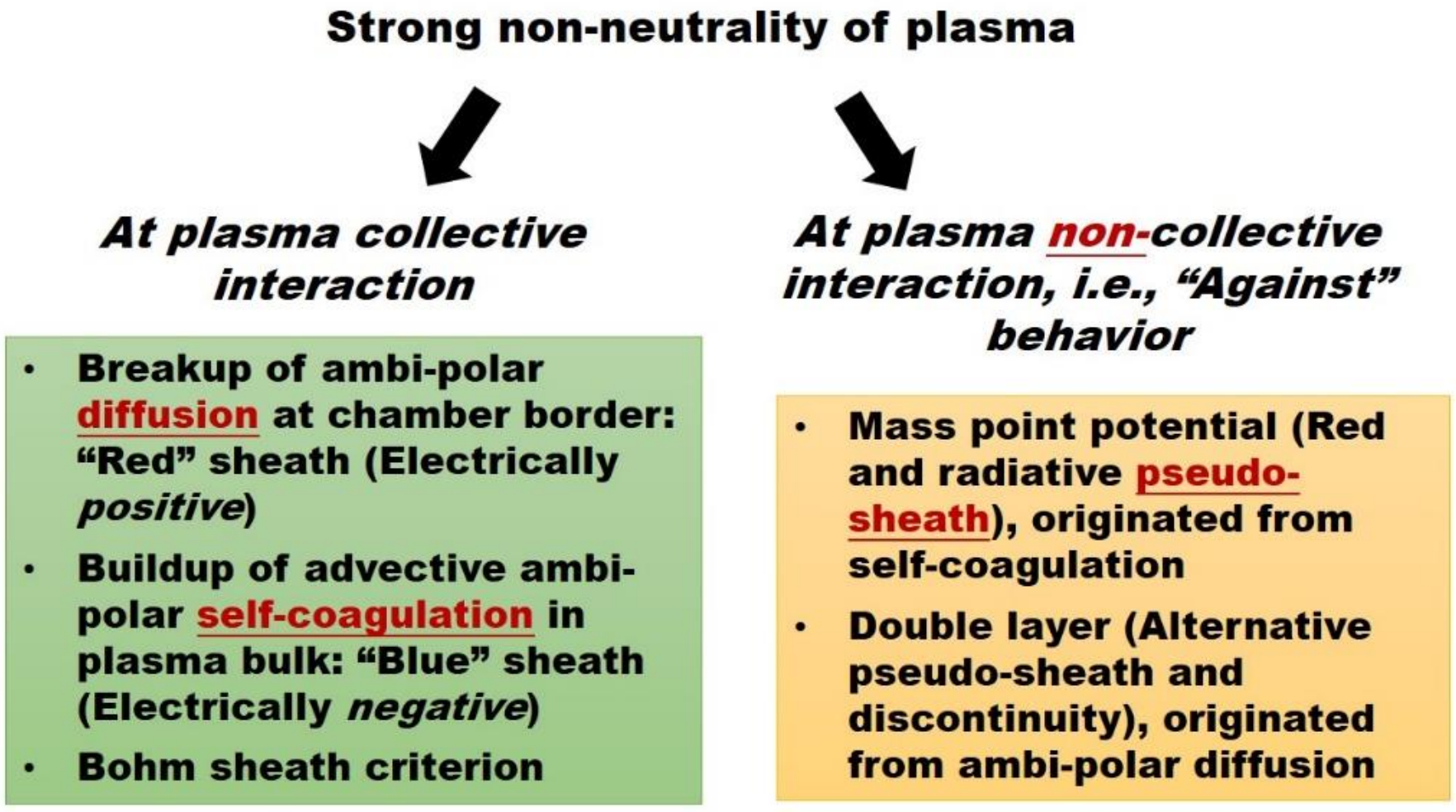

(a) For electropositive plasma and strongly electronegative plasma (*e.g.*, Ar/$SF_6$ plasma)

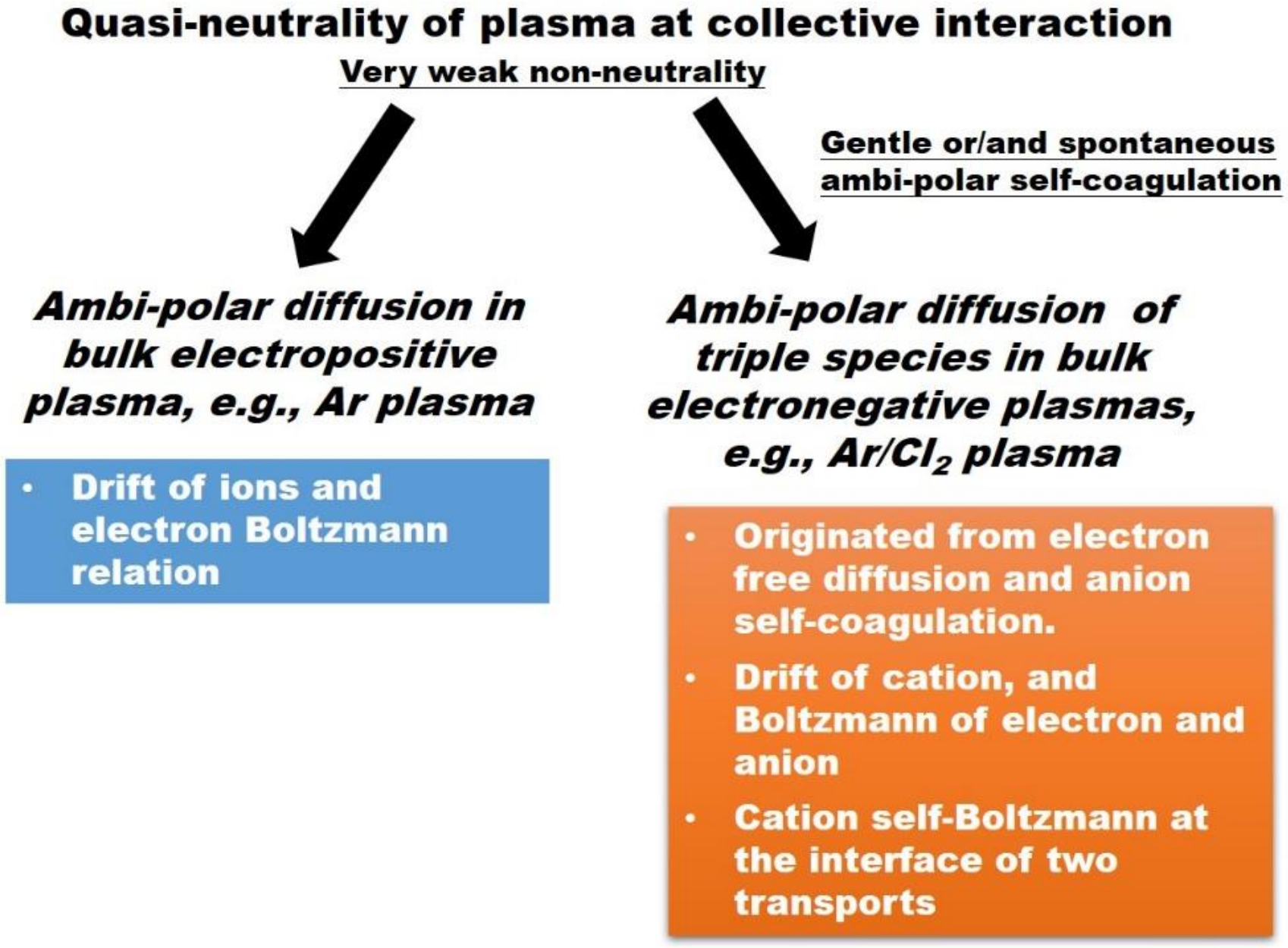

(b) For electropositive plasma and intermediate electronegative plasma

Figure 13 Schematic illustration of plasma non- and quasi- neutralities discovered presently in cold laboratory plasmas. Here, more details are provided, upon comparing to previous tabular expression.

(3.3) Decreasing the $Cl_2$ content at high pressures, *e.g.*, 90mTorr, from 90% $Cl_2$ content

(a) Collapse of self-coagulation-to-coil

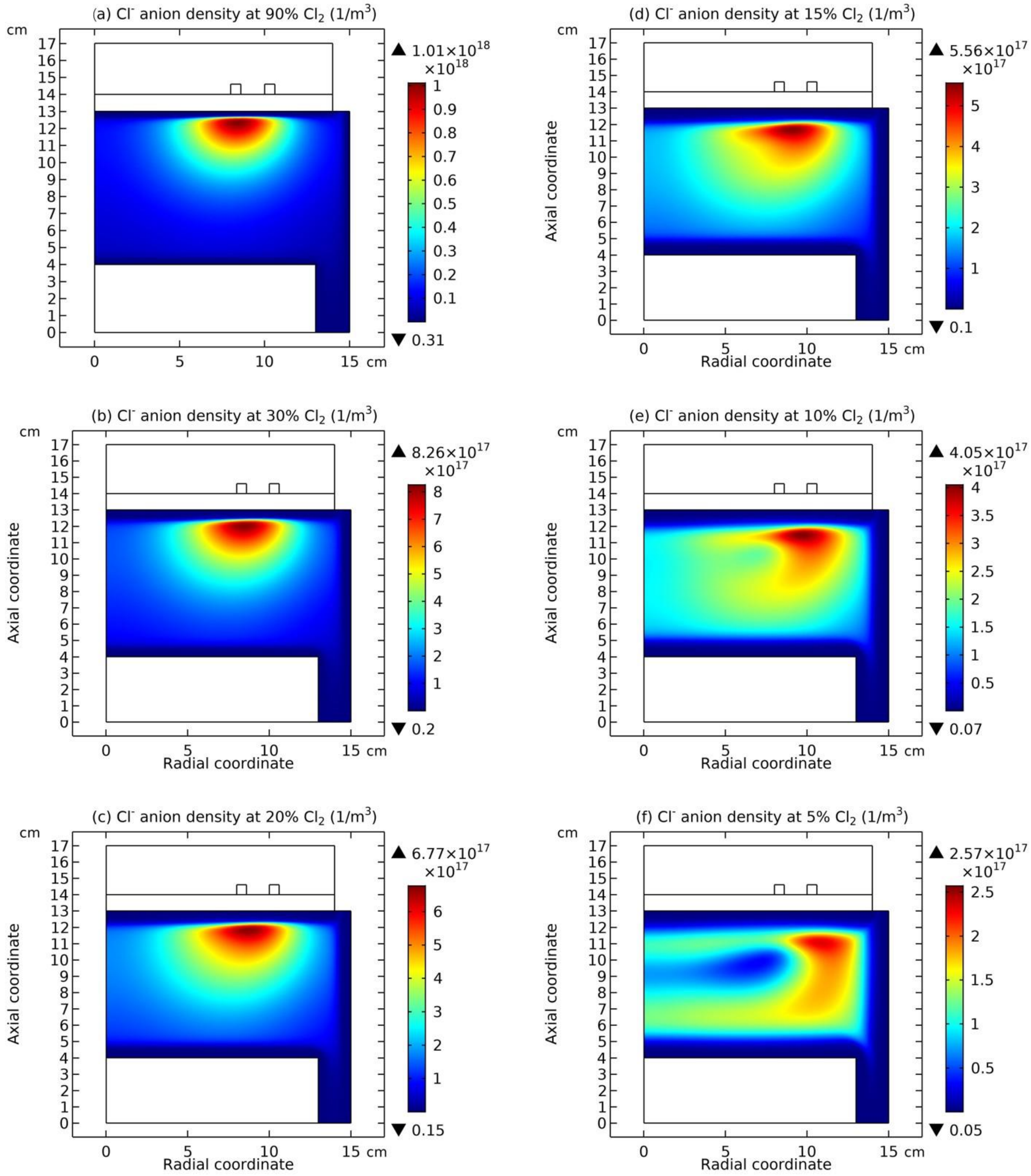

Figure 14 $Cl^-$ anion density of $Ar/Cl_2$ inductive plasma at (a) 90%, (b) 30%, (c) 20%, (d) 15%, (e) 10%, and (f) 5% $Cl_2$ contents, given by fluid simulation at 300W and 90mTorr.

In Fig. 14, the $Cl^-$ anion densities at different $Cl_2$ contents are plotted, at the high pressure of 90mTorr and 300W power of $Ar/Cl_2$ inductive plasma. It is seen that when decreasing $Cl_2$ content, the known self-coagulation-to-coil (SCC) phenomenon (reported in Ref. [16]) is disappeared gradually and hollow $Cl^-$ anion density is constructed at low enough $Cl_2$ content, *i.e.*, 5%, in Fig. 14(f).

(b) Hollow anion density and grouping effect

The hollow anion density profile is more clearly exhibited along the axis in Fig. 15. Moreover, the plasma species are grouped (novel and interesting behavior) in Fig. 16. Concretely, the electron and the cation pair of $Ar^+$ and $Cl^+$ are grouped as their density profiles are the same, coagulated to the coil. Correspondingly, the left $Cl^-$ anion and $Cl_2^+$ cation species are grouped, both displaying the hollow density profile. This grouping can be further demonstrated through the charge density in Fig. 17. In Fig. 17(a) the deduct of electron density from the pair of $Ar^+$ and $Cl^+$ is plotted and in Fig. 17(b) the deduct of $Cl^-$ density from the $Cl_2^+$ is plotted. As compared, they generate the charge densities with just opposite polarities. When summed all charged species, the final net charge density is expressed in Fig. 17(c). The basic double layer structure is more or less seen, which surrounds the weak red positive "*charge cloud*", defined as grouping potential (one more plasma non-neutrality type).

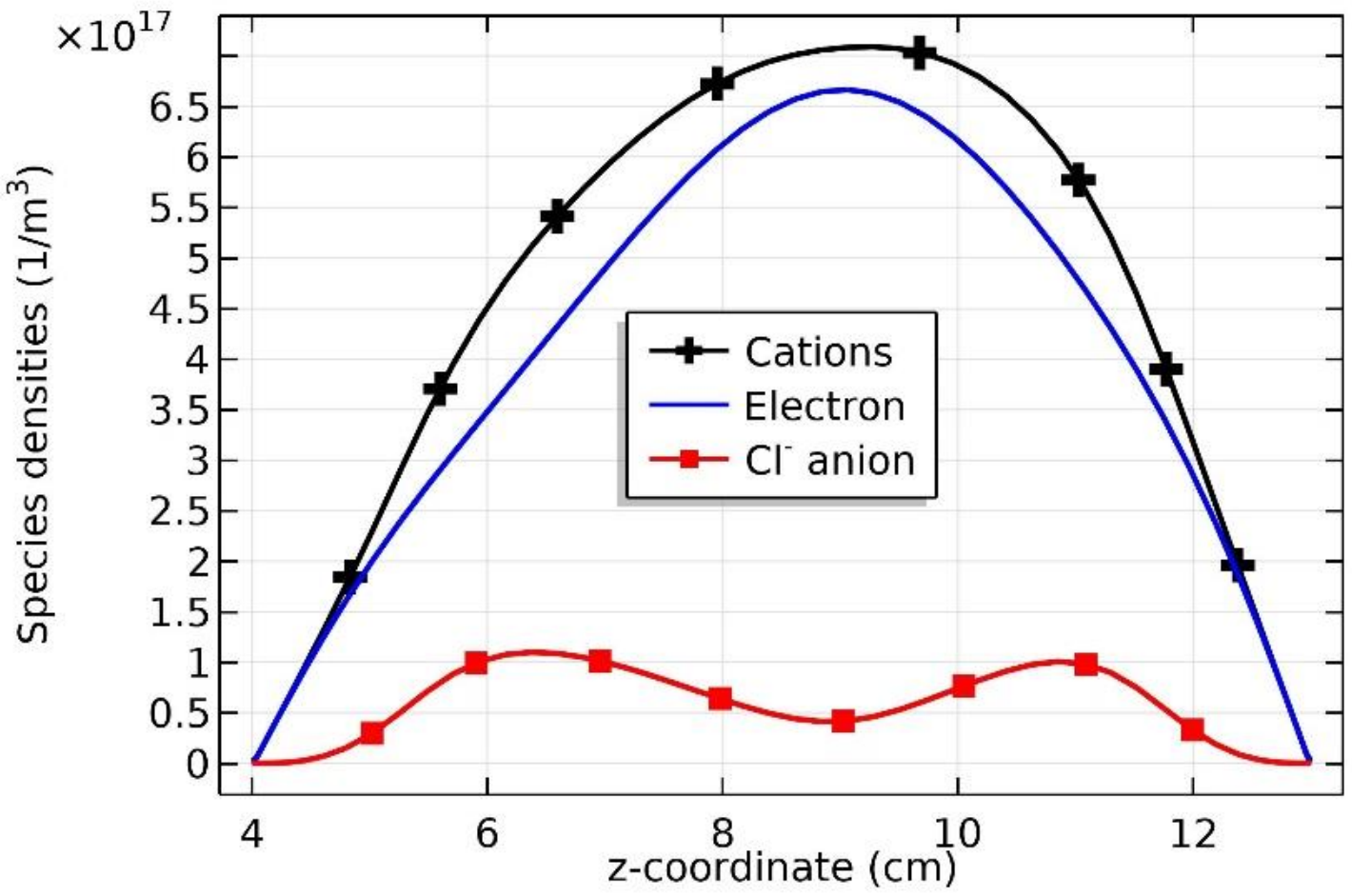

Figure 15 Axial profiles of summed cations, electron and $Cl^-$ anion densities of $Ar/Cl_2$ inductive plasma, at 5% $Cl_2$ content and 90mTorr, and 300W, simulated by fluid model. Herein, the hollow anion density is shown.

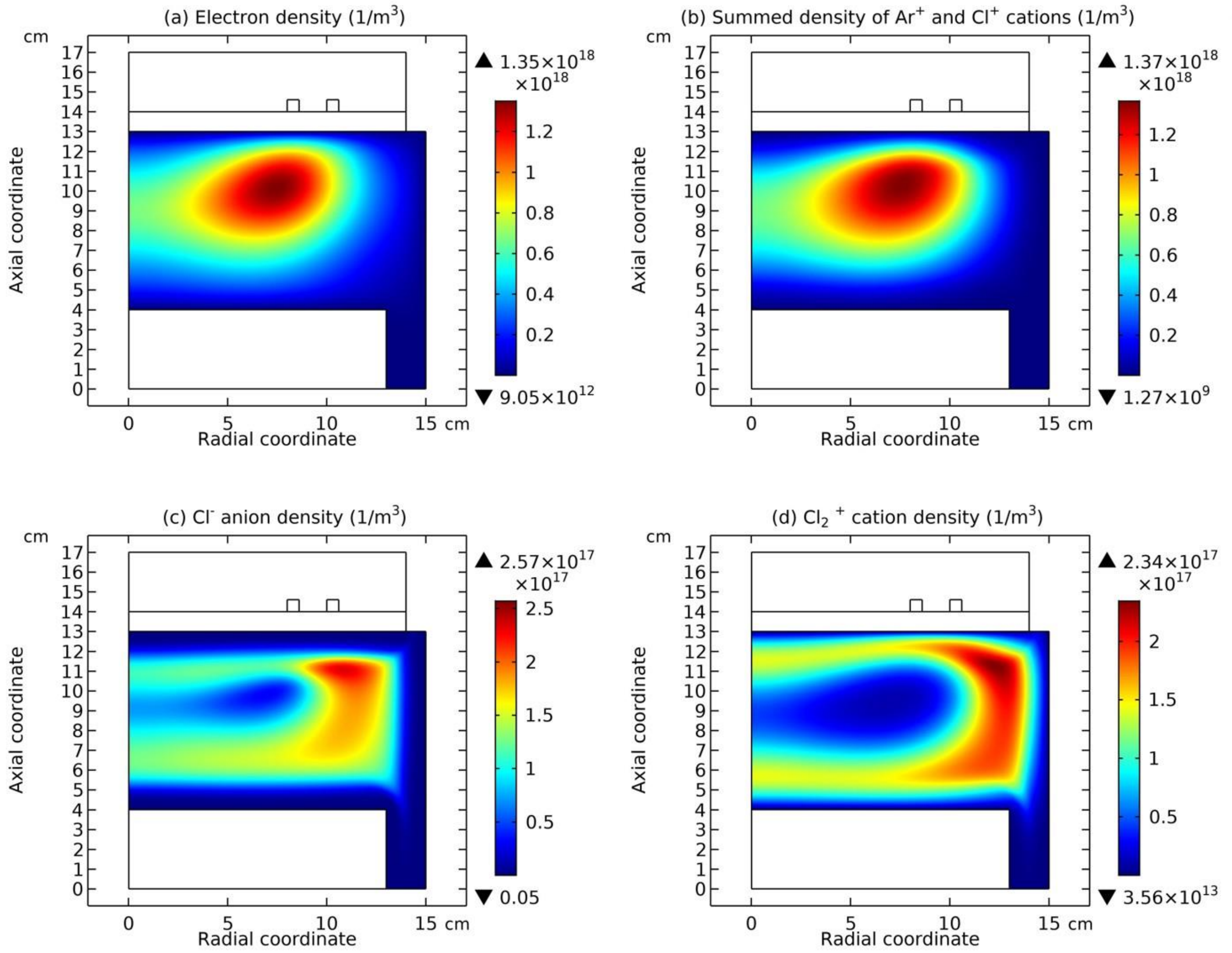

Figure 16 (a) Electron density, (b) summed $Ar^+$ and $Cl^+$ cations density, (c) $Cl^-$ anion density, and (d) $Cl_2^+$ cation density of $Ar/Cl_2$ inductive plasma at 5% $Cl_2$ content, 90mTorr and 300W, given by fluid simulation. Herein, the grouping effect is shown, where the coagulated electron and pair of $Ar^+$ and $Cl^+$ are grouped and individual hollow $Cl^-$ and $Cl_2^+$ are grouped.

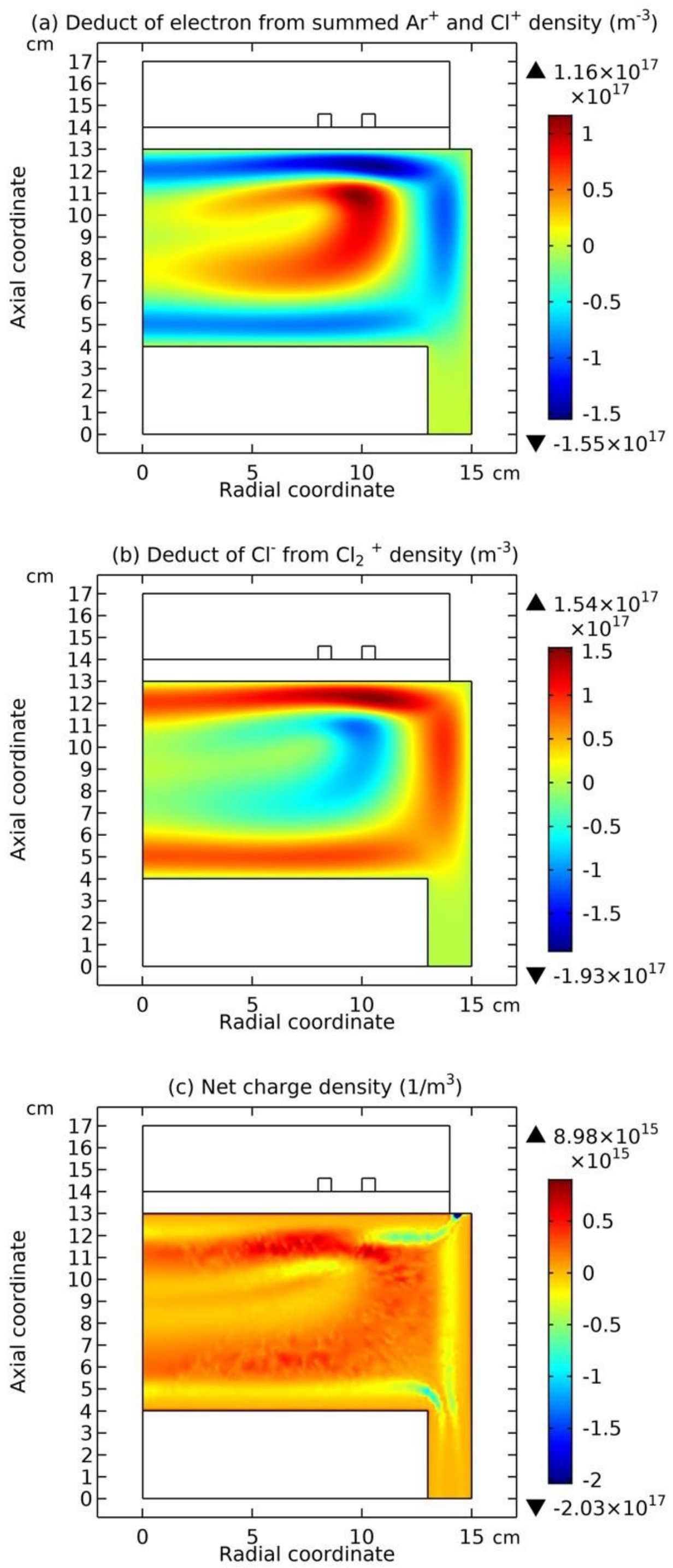

Figure 17 (a) Deduct of electron from the $Ar^{+}$ and $Cl^{+}$ pair density, (b) deduct of $Cl^{-}$ anion from $Cl_2^{+}$ cation density, and (c) net charge density of Ar/$Cl_2$ inductive plasma at 5% $Cl_2$ content, 90mTorr and 300W, given by fluid simulation. Herein, the grouping effect is shown more clearly, where the opposite charges are given by the two groups, for basically satisfying the electric neutrality of plasma.

## (c) Reactive feedstock gas depletion and shortage

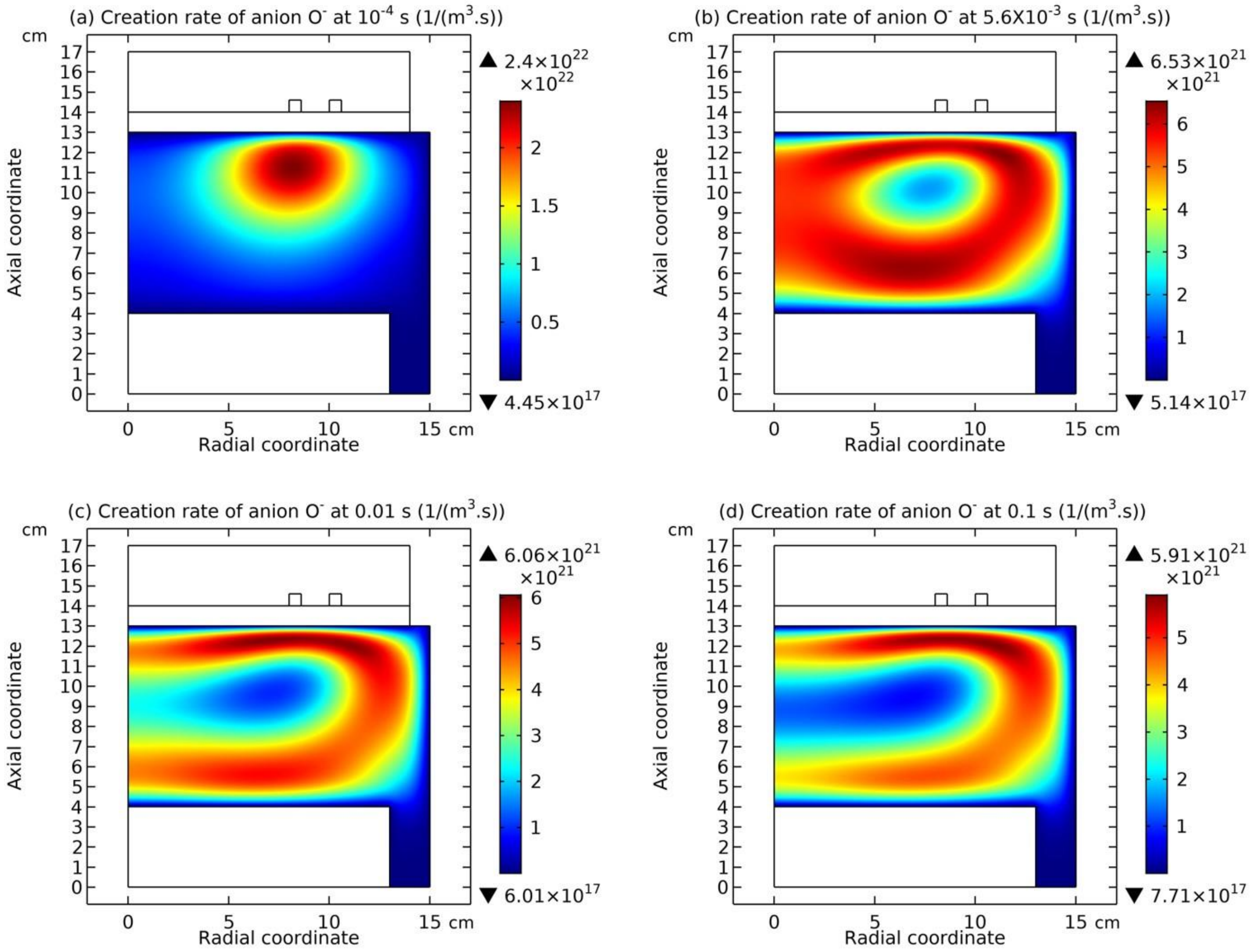

Figure 18 Creation rate of $O^-$ anion at different times, (a) $10^{-4}$s, (b) $5.6\times10^{-3}$s, (c) $0.01$s and (d) $0.1$s, in the $Ar/Cl_2$ inductive plasma given by fluid simulation at 300W, 5% $Cl_2$ content and 90mTorr.

Why the hollow anion density is happened when decreasing $Cl_2$ content? Because of the shortage of feedstock reactive gas, $Cl_2$ molecule. As shown in Fig. 18, at 5% $Cl_2$ and 90mTorr, as increasing the simulating time in fluid model simulation, the creation rate of $Cl^-$ anion, *i.e.*, attachments of electron with $Cl_2$, at first peaks under the coil and then sinks, finally exhibiting centric hollow shape, implying that the feedstock $Cl_2$ molecules have been severely depleted. In Fig. 19, the $Cl_2$ density at the different times of Fig. 18 are plotted. As seen, the feedstock is indeed consumed because of the small $Cl_2$ content in the feeding gas mixture.

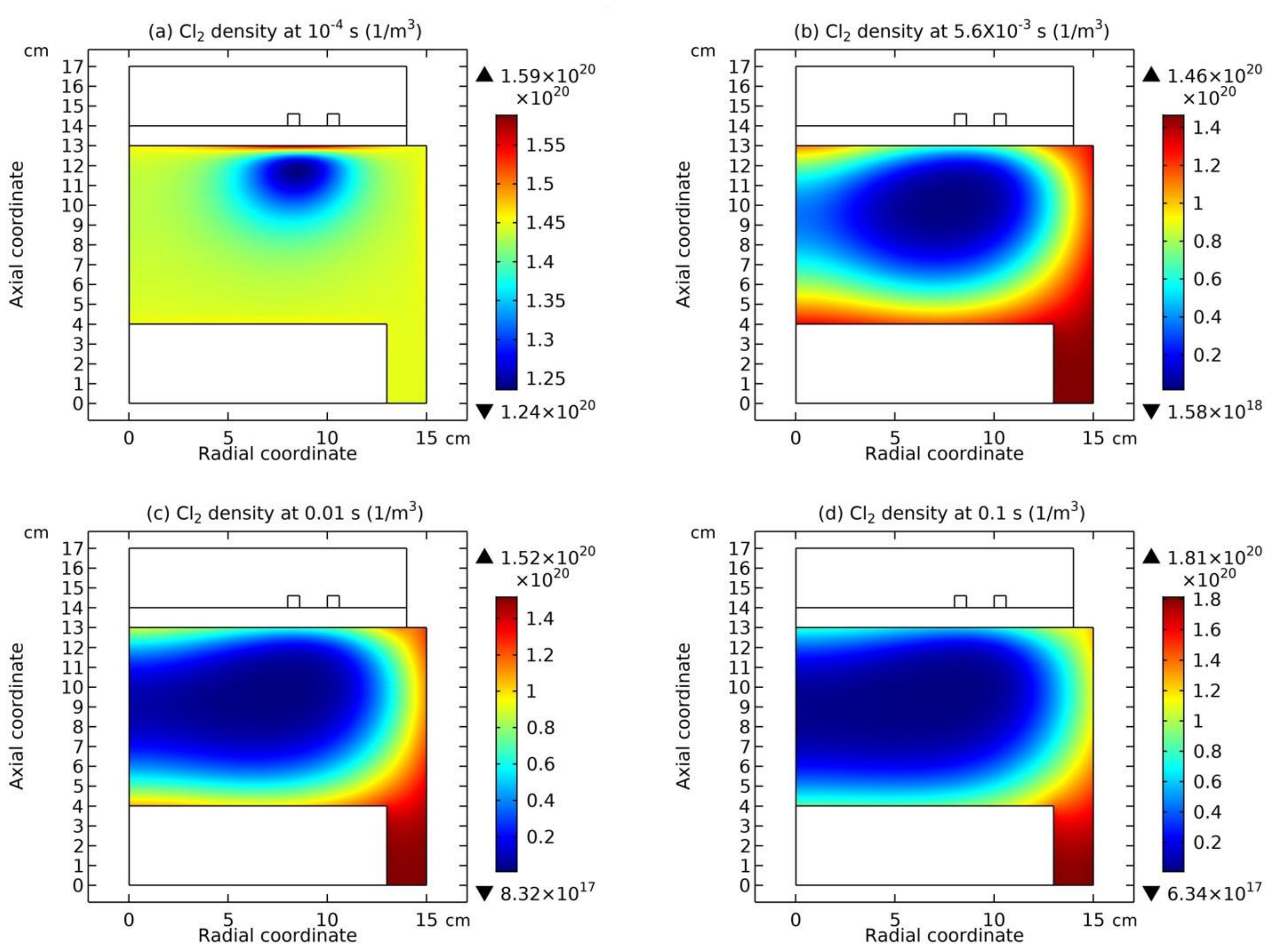

Figure 19 Density of one feedstock gas, $Cl_2$, of the Ar/$Cl_2$ inductive plasma at different times, (a) $10^{-4}$s, (b) $5.6\times10^{-3}$s, (c) $0.01$s and (d) $0.1$s, given by the fluid simulation at the discharge conditions of 300W, 5% $Cl_2$ content and 90mTorr.

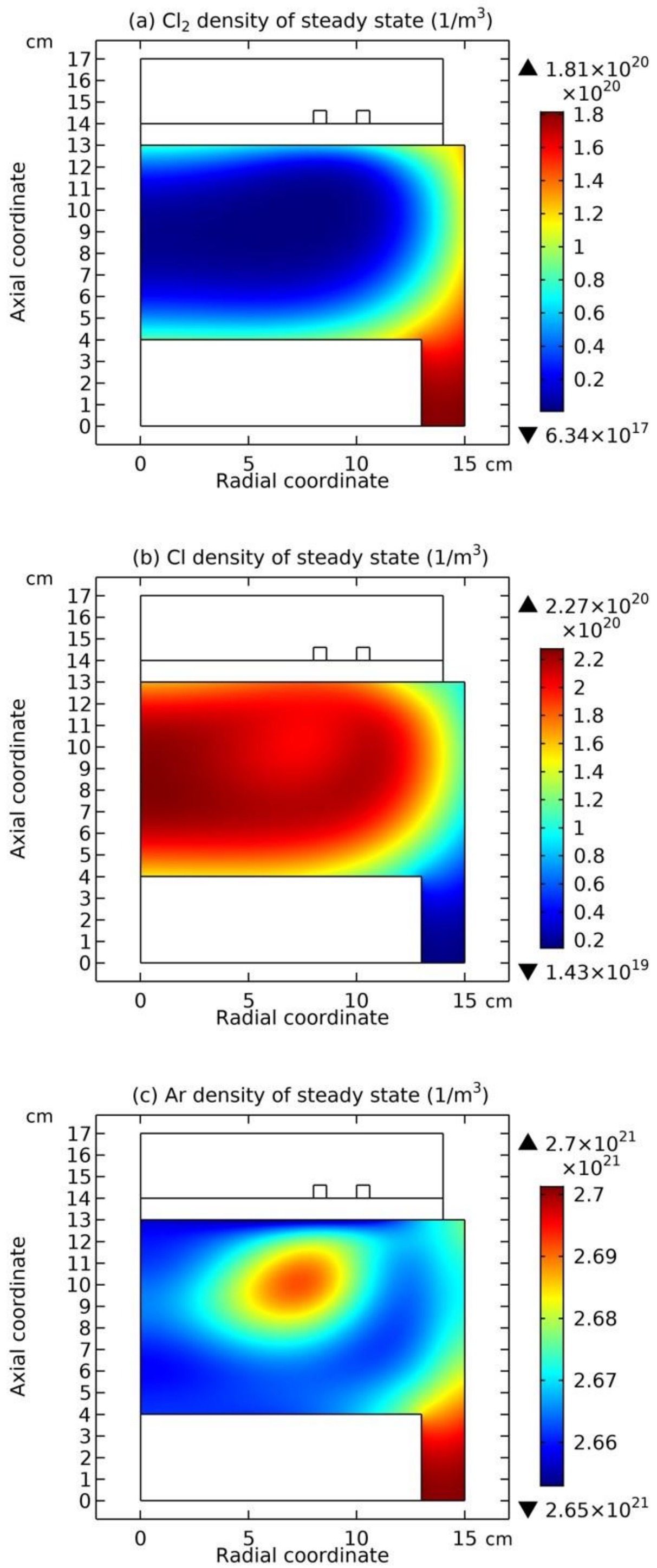

Figure 20 Densities of $Cl_2$ (a), Cl (b) and Ar (c) neutrals of steady state fluid model simulation in the Ar/$Cl_2$ inductive plasma at 300W, 5% $Cl_2$ content and 90mTorr.

In Fig. 20, the densities of neutrals, $Cl_2$, Cl and Ar, at 5% $Cl_2$ and 90mTorr are plotted. In Fig. 21, the axial profiles of these three densities under the dielectric window are given. It is seen that the Ar and Cl densities are high and the $Cl_2$ density is about two or three orders lower than the Ar and Cl at the discharge center. This explains well the grouping effect, for Ar and Cl mainly generate $Ar^+$ and $Cl^+$, while $Cl_2$ gives rise to $Cl_2^+$ and $Cl^-$, from the point view of chemistry. That's why the electron (always at enough feedstock) and the pair of $Ar^+$ and $Cl^+$ cations are grouped, and $Cl^-$ and $Cl_2^+$ are grouped. The latter is in the condition of feedstock gas shortage while the former is not.

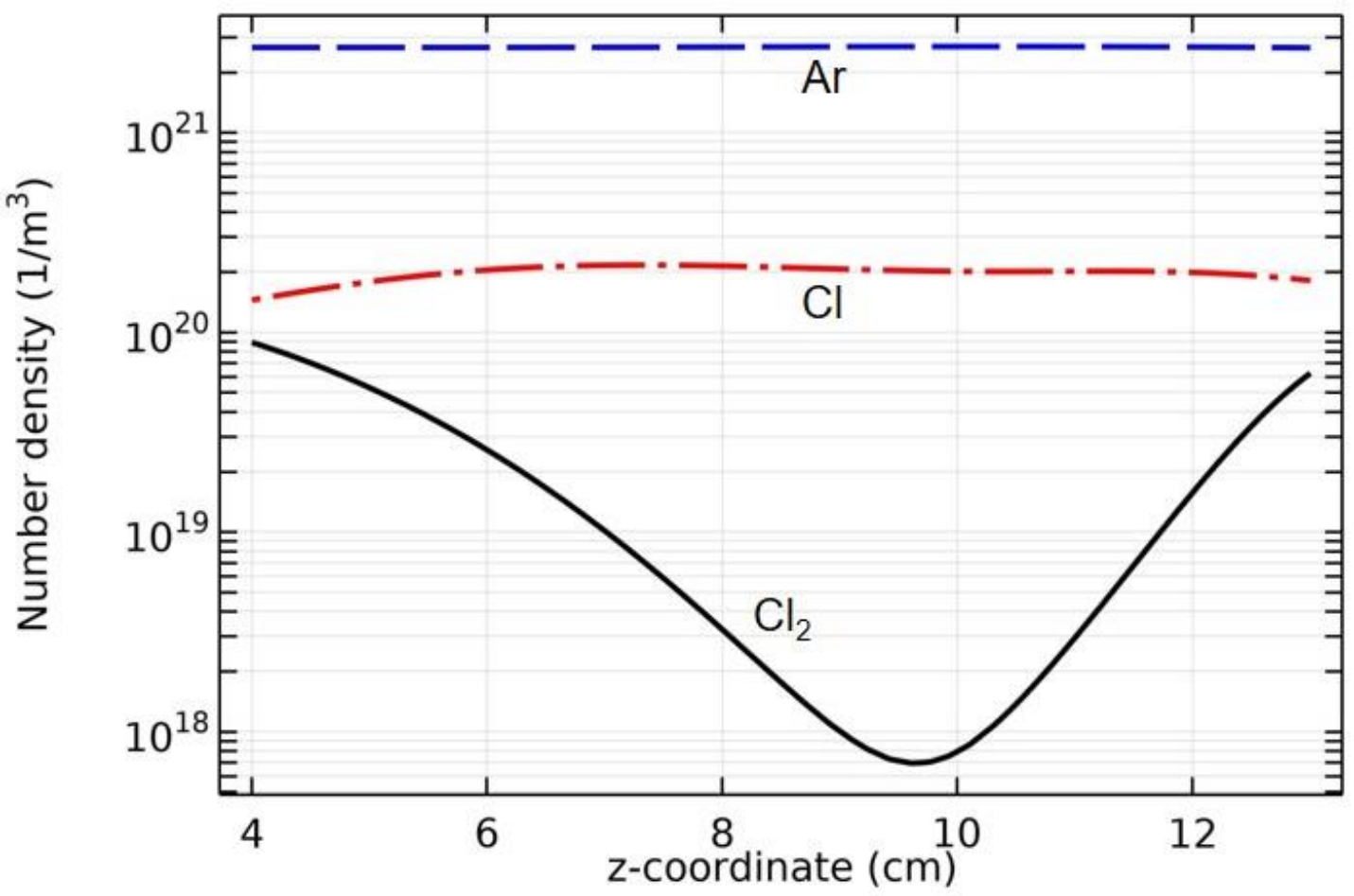

Figure 21 Axial profiles of $Cl_2$, Cl and Ar neutrals densities of steady state fluid simulation of Ar/$Cl_2$ inductive plasma at 300W, 5% $Cl_2$ content and 90mTorr. The radial location is selected under the dielectric window.

### (d) Electron coagulates at ambi-polar diffusion, without any self-coagulation

In Fig. 22, the electron density and potential contour of 5% $Cl_2$ and 90mTorr Ar/$Cl_2$ plasma are given. As seen, when the self-coagulation-to-coil scheme is destroyed by the shortage of reactive feedstock, the electron coagulation shape is different. It does not attach the dielectric window anymore without the influence of ions self-coagulation. This natural coagulation of electron is one general electropositive plasma characteristic. It originates from the fact that the electron ionization frequency of chemical term cannot be taken as constant in the electron continuity equation at high pressure, which is originated from the non-uniformity of electron temperature. Hence, the trigonometric or Bessel profile (*i.e.*, the analytic solutions of continuity equation at constant frequency [5]) is not given, but forming the coagulated one. This is thus called the physical coagulation, relative to the chemical self-coagulation of ions (in electronegative core). Besides, the electron Boltzmann balance with plasma potential is automatically met as the peripheric self-coagulation of electron is neither happened herein (strong potential barrel at weak electronegativity).

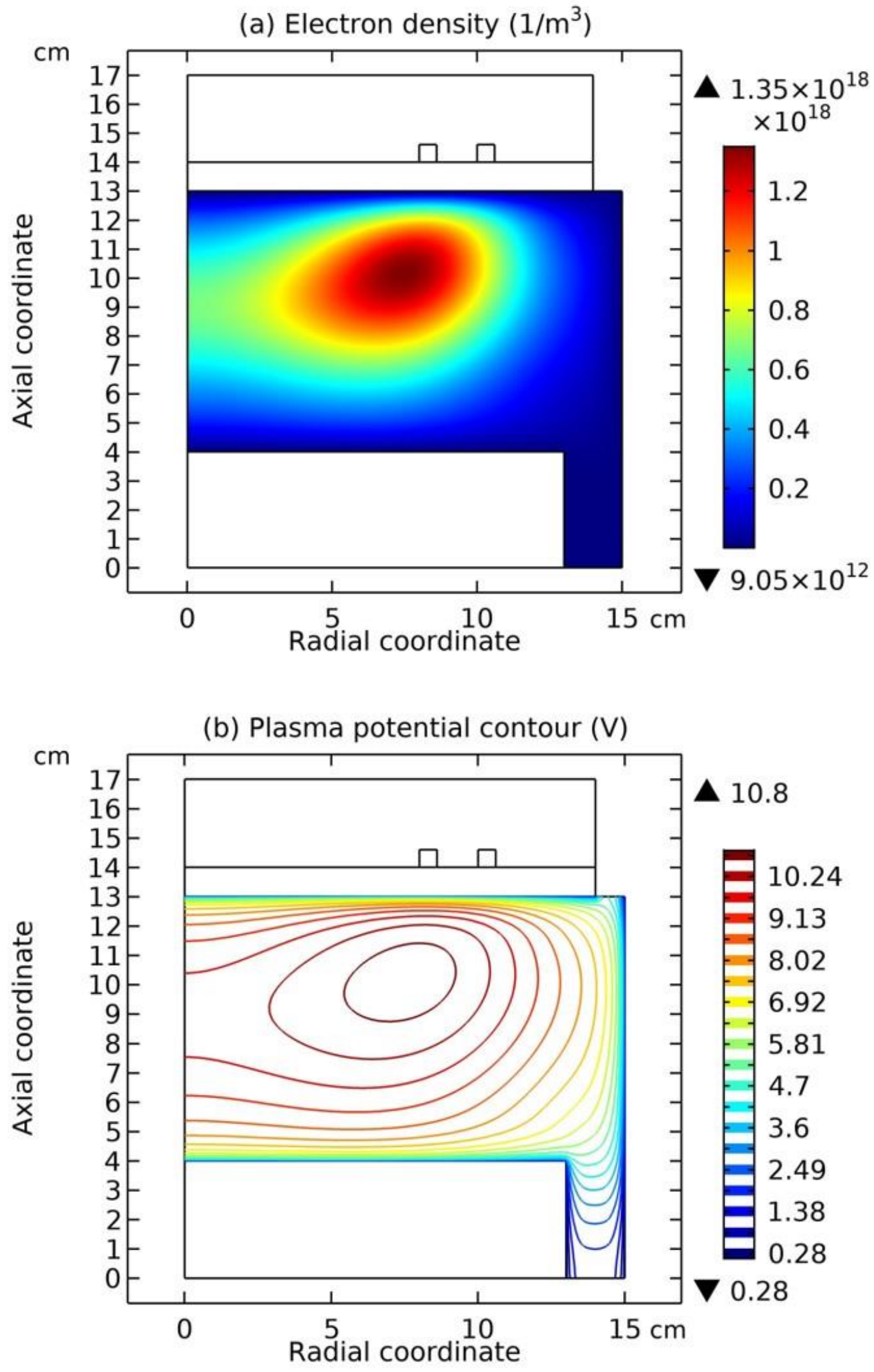

Figure 22 Electron density (a) and plasma potential (b) of $Ar/Cl_2$ inductive plasma, given by fluid simulation at the discharge conditions of 300W, 5% $Cl_2$ content and 90mTorr.

(3.4) Decreasing the pressure at low $Cl_2$ contents, *e.g.*, 5%, from 90mTorr

(a) From hollow to $\delta$ type and disappearance of grouping

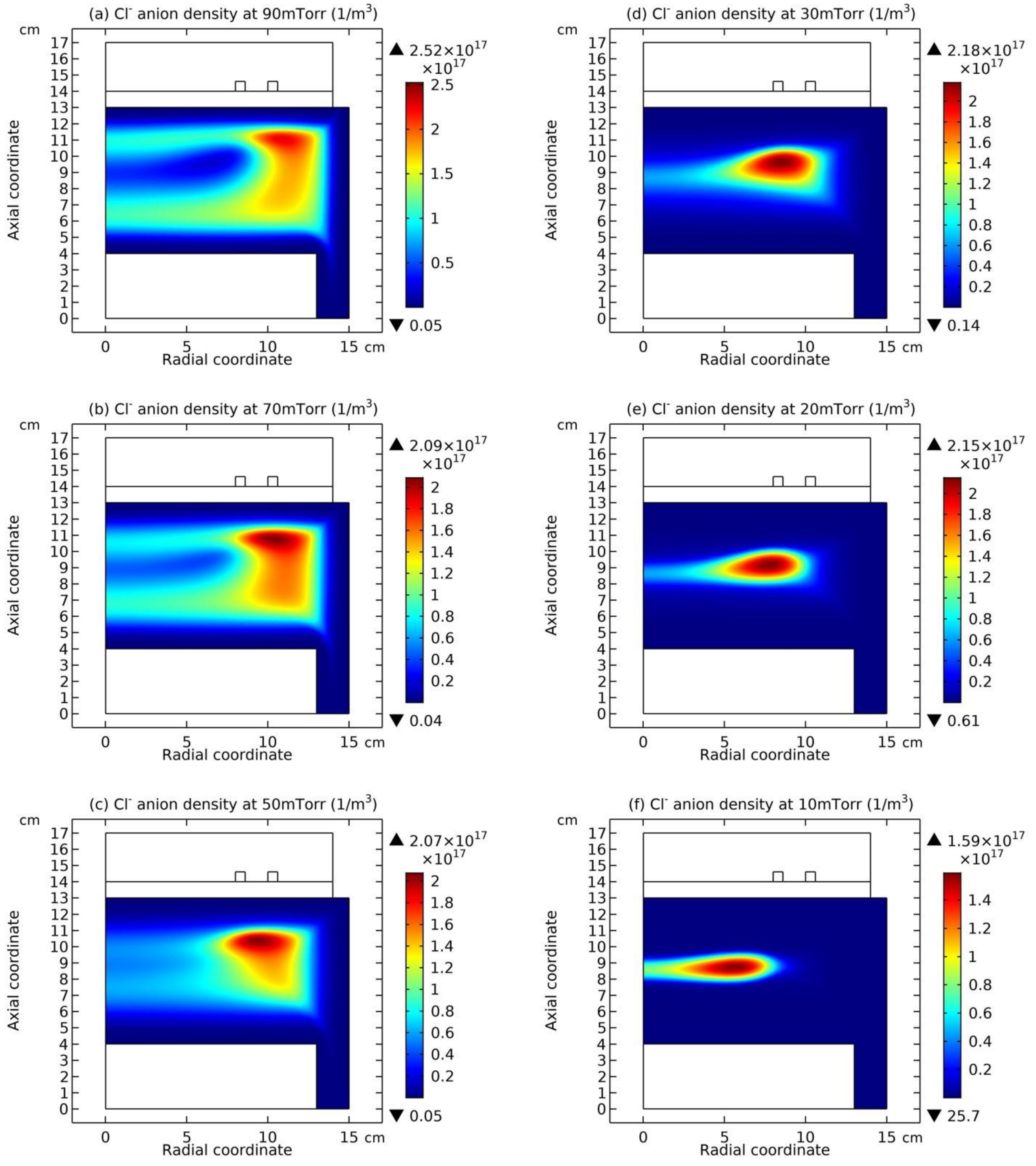

Figure 23 $Cl^-$ anion density of $Ar/Cl_2$ inductive plasma at different pressures, (a) 90mTorr, (b) 70mTor, (c) 50mTorr, (d) 30mTorr, (e) 20mTorr and (f) 10mTorr, given by the fluid model simulation at the discharge conditions of 300W, 5% $Cl_2$ content.

In Fig. 23, when decreasing pressure at low $Cl_2$ content, 5%, the hollow anion density is disappeared and the delta type anion is reoccurred at 10mTorr. Furthermore, the grouping effect is disappeared as well. As shown in Fig. 24, at 10mTorr and 5% $Cl_2$, all cations densities hold the delta part as anion, at the advective ambi-polar self-coagulation.

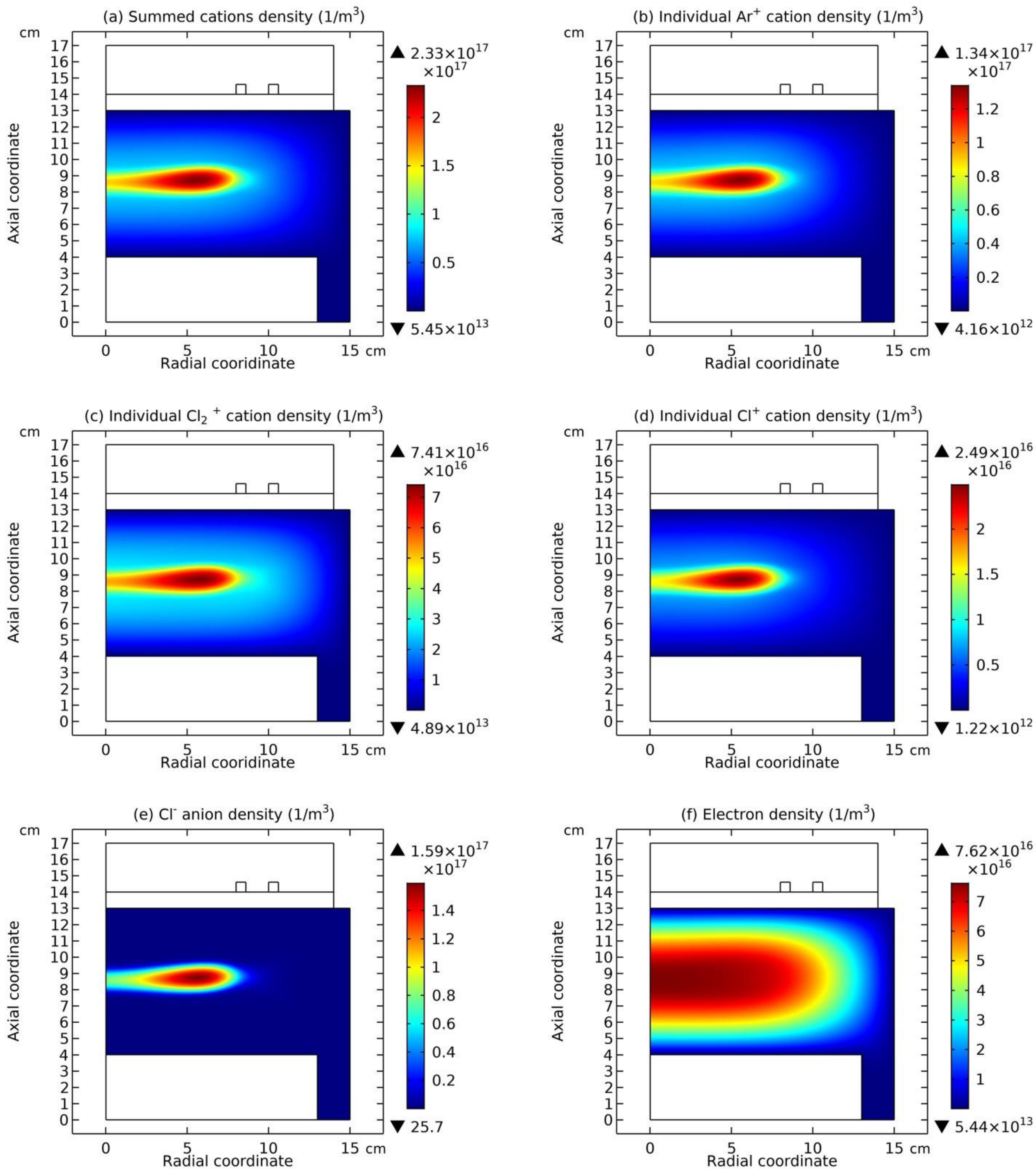

Figure 24 Summed cations density (a), the individual $Ar^+$ (b), $Cl_2^+$ (c) and $Cl^+$ (d) densities, $Cl^-$ anion density (e) and the electron density (f) of $Ar/Cl_2$ inductive plasma, given by the fluid model simulation at the discharge conditions of 300W, 5% $Cl_2$ content and 10mTorr.

(b) Refreshment of reactive feedstock gas and de-coagulation of electron

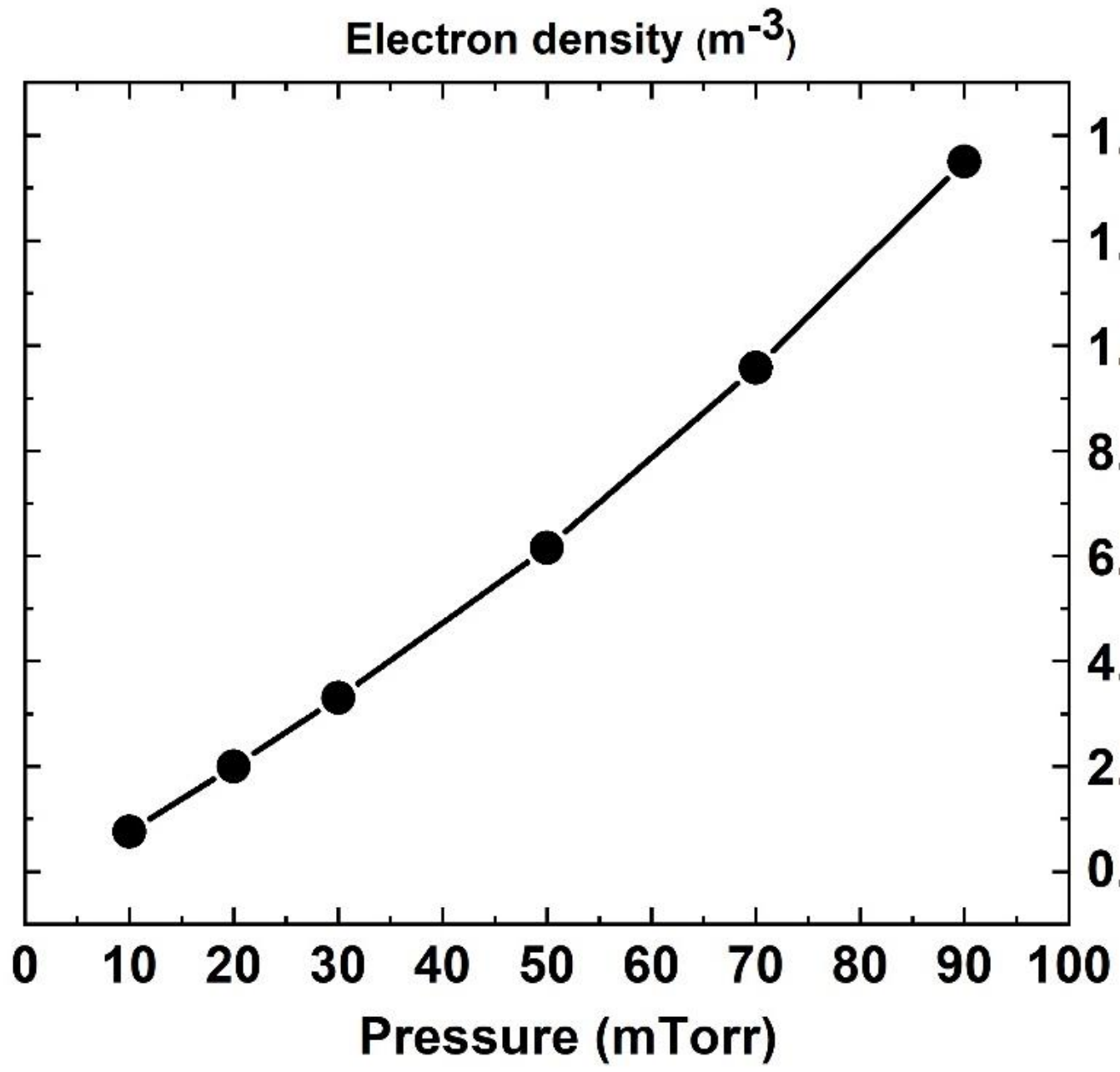

Figure 25 Electron density peak versus gas pressure, at 300W and 5% $Cl_2$ content, at fluid simulation.

The disappearance of hollow anion and group effect is because electron density is low at low pressure in Fig. 25, which consumes less feedstock gases. So, the reactive feedstock gas is refreshed. Hence, the creation rate of anion is not centrically sunk at decreasing the pressure, as shown in Fig. 26. In Fig. 27, the Ar, Cl and $Cl_2$ densities axial profiles of 10mTorr and 5% $Cl_2$ are plotted. It is seen that the $Cl_2$ feedstock is just two times lower than the Ar and Cl, illustrating the group disappearance (for all charges species are now in enough feedstocks). In Fig. 28, it is shown that the electron coagulation is disappeared and it turns back to the trigonometric (along axis) and basic Bessel (along radius) profiles on decreasing the pressure at low $Cl_2$ content. This is logic, since at low pressure the transport (with larger diffusion coefficient and mobility than the high pressure) dominates over the chemical term, and the spatial variation of ionization frequency is negligible in the electron continuity equation. It is noted that due to the influence of delta anion, the radial profile of electron density herein is slightly deviated from the Bessel function, just as illustrated in our previous $Ar/O_2$ inductive plasma [17].

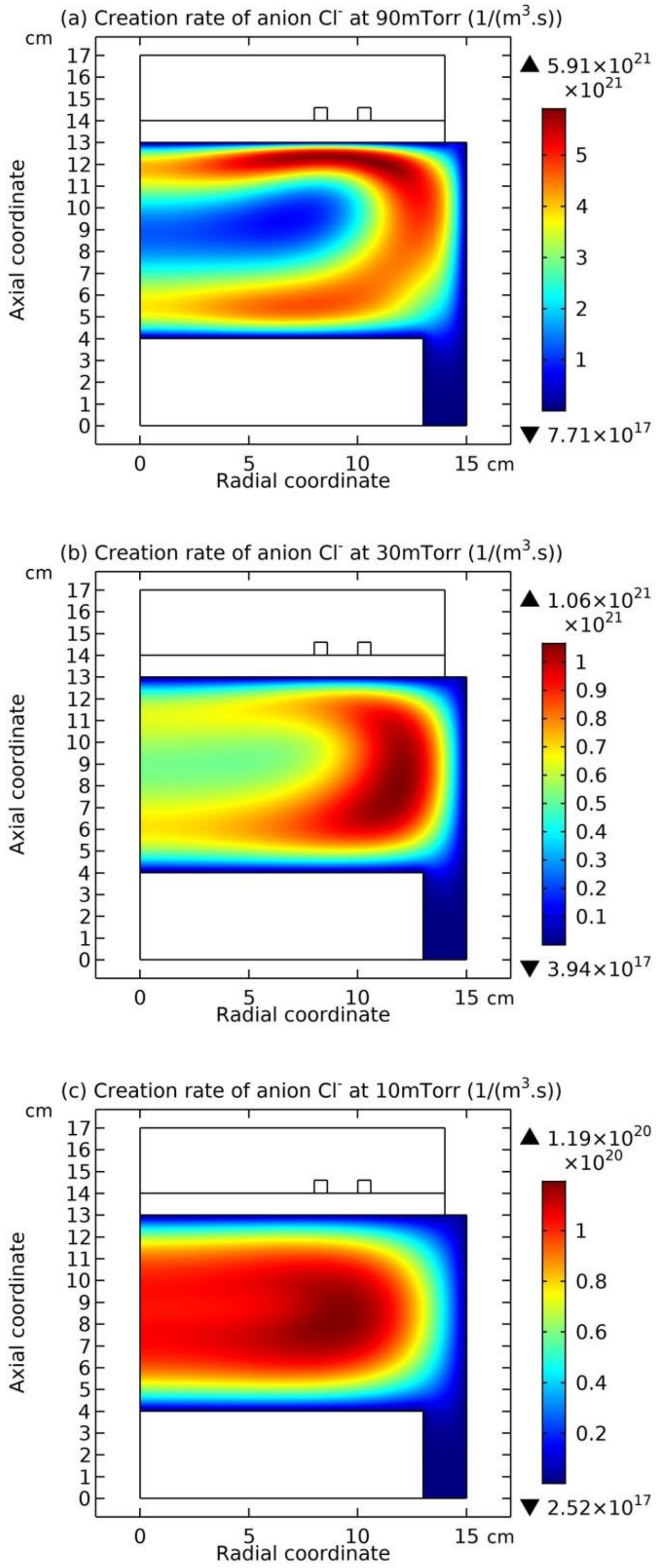

Figure 26 Creation rates of $Cl^-$ anion at different pressures, (a) 90mTorr, (b) 30mTorr and (c) 10mTorr, of $Ar/Cl_2$ inductive plasma given by the fluid model simulation at 300W, 5% $Cl_2$ content.

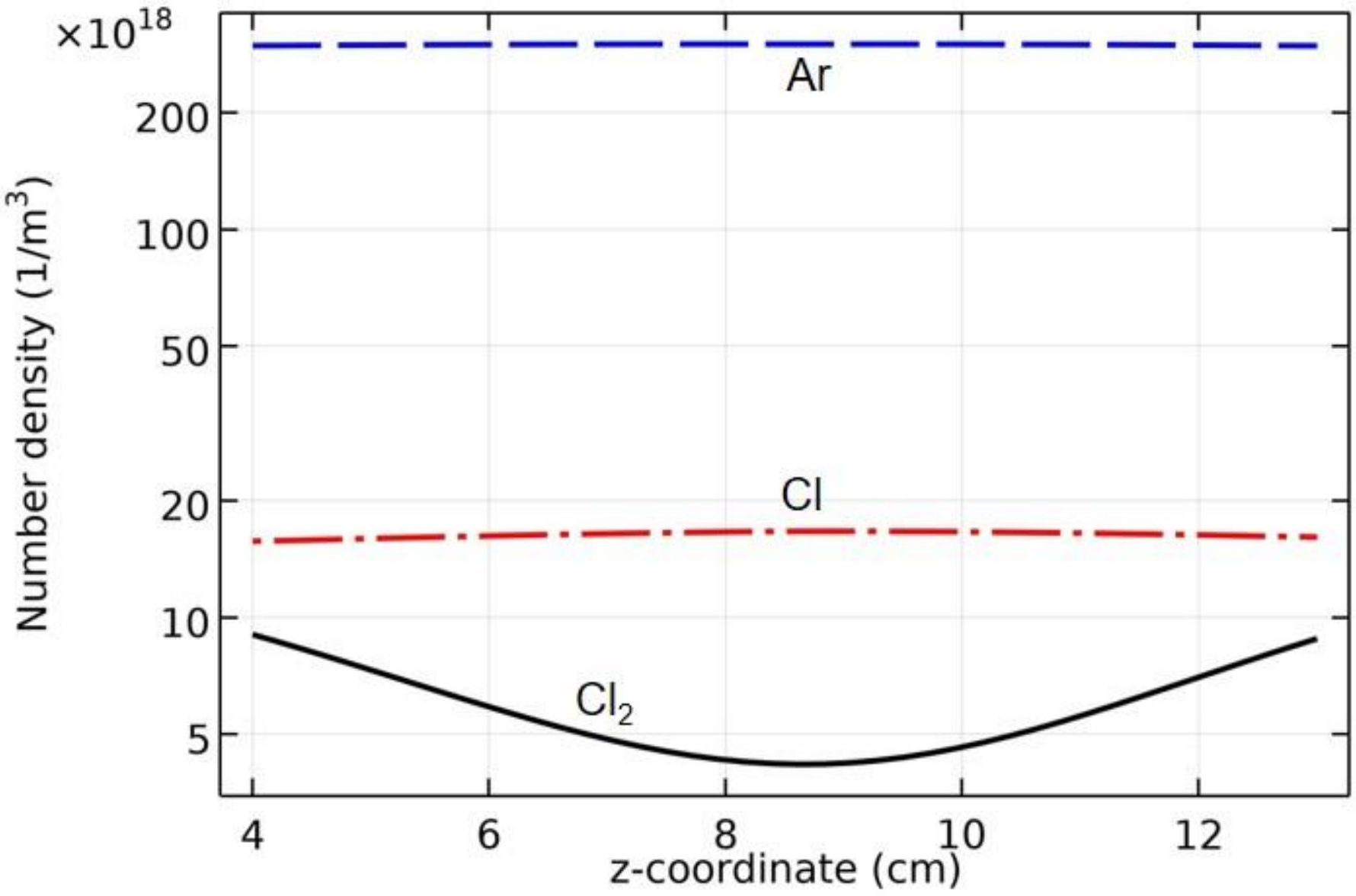

Figure 27 Axial profiles of Ar, Cl and $Cl_2$ neutrals densities of Ar/$Cl_2$ inductive plasma given by the fluid model simulation at 300W, 5% $Cl_2$ content and 10mTorr, under the dielectric window.

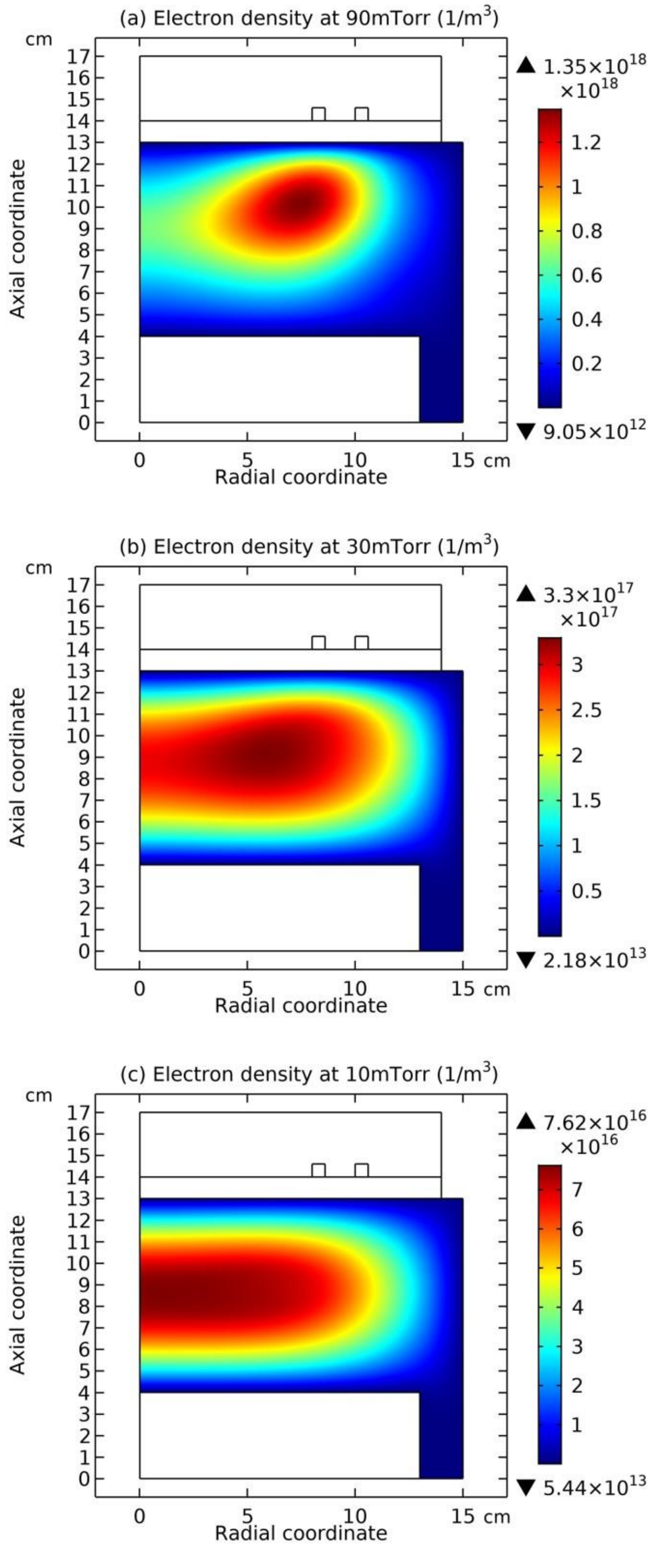

Figure 28 Electron density profile at different pressures, (a) 90mTorr, (b) 30mTorr and (c) 10mTorr, given by the fluid model simulation at 300W and 5% $Cl_2$ content. In this plot, the de-coagulation of electron is demonstrated.

## IV. Conclusion and further remarks

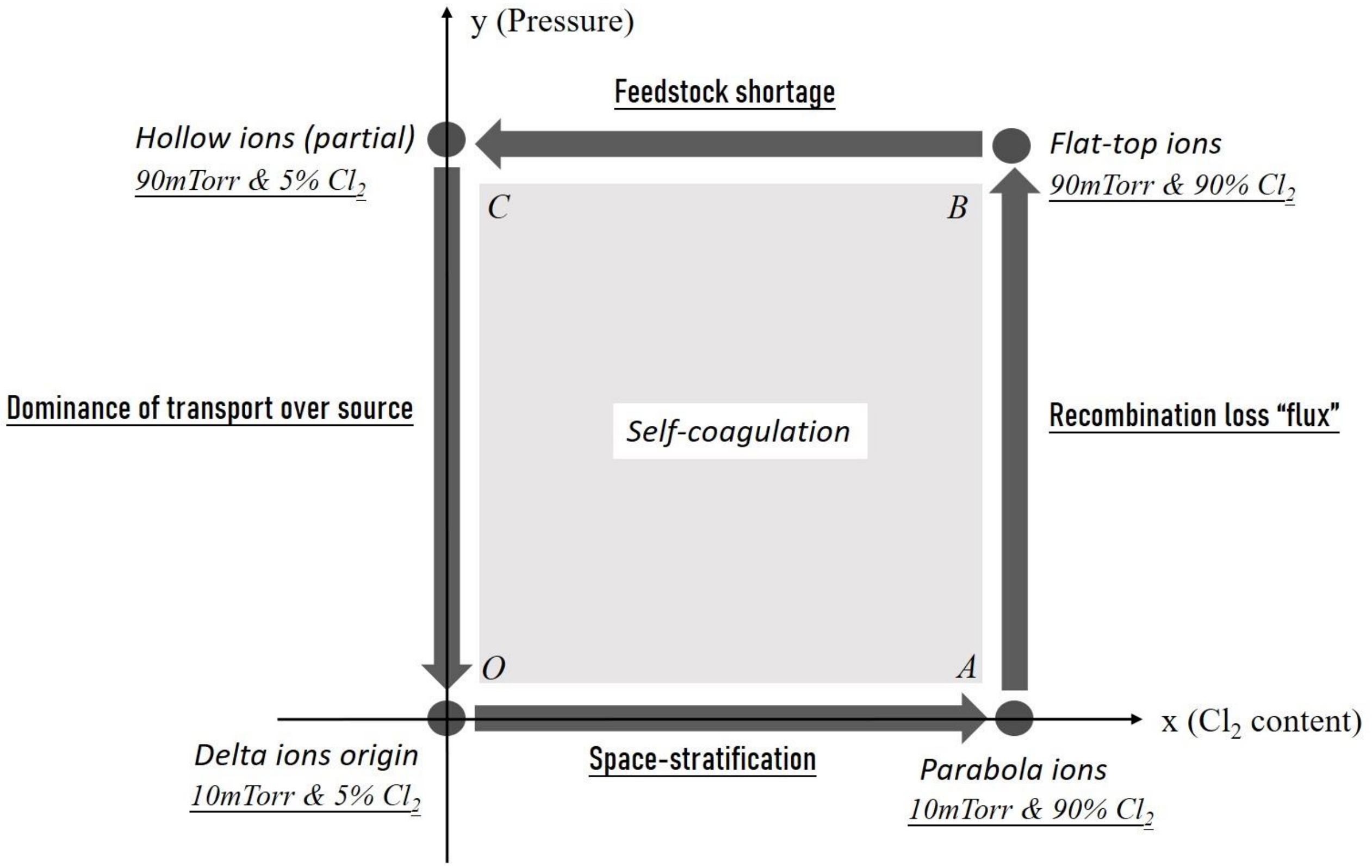

Figure 29 (Partial) overview of the cyclic study with respect to discharge conditions, *i.e.*, pressure and reactive gas content concentration, at fixed power 300W and in the inductive Ar/$Cl_2$ plasma given by fluid model simulation.

Partial contents of the cyclic study involving the pressure and $Cl_2$ content parameters of inductive Ar/$Cl_2$ plasma at fixed power (300W) are schematically illustrated in one transformed Cartesian coordinate system, as shown in Fig. 29. The x axis represents the $Cl_2$ content, and y axis represents pressure, respectively. The origin of this system is about the delta anion, which is formed at 10mTorr and 5% $Cl_2$ content. Anion self-coagulation produces its own delta and cation delta is led to by the ambi-polar self-coagulation. So, the delta ions (here referring to both anion and cation) term is appeared at the origin. At the center of drawn rectangle, the word, *self-coagulation*, is written, implying its importance and influence in almost all the processes occurring in the cycle. At increasing the $Cl_2$ content (along x axis) from the delta origin (O vertex of rectangle), after experiencing the space-stratification and core expansion, the parabola ions model is obtained, at 10mTorr and 90% $Cl_2$ content (vertex A). When increasing the pressure from the parabola ions, at the influence of so-called recombination loss flux, parabola turns into flat-topped model at 90mTorr and 90% $Cl_2$ content (vertex B). When decreasing the $Cl_2$ content at high pressure, the reactive feedstock gas shortage is met, and the partial hollow ions model is predicted, at 90mTorr and 5% $Cl_2$ content (vertex C). Eventually, when decreasing pressure (along y axis) back to 10mTorr at 5% $Cl_2$, the transport dominates over source term, and the delta ions model is re-obtained. This constructed rectangle describes the processes occurred in the cycle that are easily edited using this coordinate method (that's why a "partial" word is added in front of the overview in the caption of Fig. 29). It is stressed there are still many other important mechanisms, *e.g.*, species-stratification (predicted by the analytics and repeated by simulation) hidden in the delta, concept of the gentle ambi-polar self-coagulation, drift essence of cation in the ambi-polar diffusion-self-coagulation, collapse of the self-coagulation-to-coil scheme, discharge grouping effect, as well as the chemical (self-) and physical (ionization frequency is not constant) coagulations of electron and its de-coagulation. Besides, the diverse non-neutralities of plasma discovered at present are summarized, analyzed, and classified, which helps people greatly understand the complex plasma. All these contents comprise this article.

Via the self-consistent fluid simulations, this article, together with its twin paper [16] mainly about the $Ar/SF_6$ inductive discharge, predicts so many new phenomena occurred in electronegative plasmas. From the tightly correlative analytic theories, all the new behaviors are well interpreted, reliable to readers. However, the validation of them through experiments is still needed, satisfying develop rules of scientific knowledges. This becomes our next work emphasis. Besides, the influences of other fluid model factors, such as inertia term, neutral gas heating and convection, and elaborate chemical reactions, on these mechanisms are also very meaningful tasks, deserving careful considerations.

## Acknowledgements

This work is financially supported by the foundation of DUT19LK59, one fundamental research fund of central universities of China. Besides, the undergraduate student, Jin-Shuo Zhang, is thanked because of his study on revealing the transport essence of inductively coupled argon plasma via fluid model simulation (supervised by the author of Shu-Xia Zhao, as one teacher), when he was attending the National Undergraduate Plasma Scientific and Technological Innovation Contest of China. This study of him helped us understand the physics coagulation (at high pressure) and de-coagulation (at low pressure) of electron in the $Ar/Cl_2$ inductive plasma at very low $Cl_2$ content.

## Conflict of interest

The authors have no conflicts to disclose.

## Data available statement

The data that support the findings of this study are available within the article.

## References

[1] P A Miller, G A Hebner, K E Greenberg, P D Pochan, and B P Aragon, 1995 *J. Res. Natl. Inst. Stand. Technol.* **100** 427.

[2] S Mouchtouris and G Kokkoris, 2016 *Plasma Sources Sci. Technol.* **25** 025007.

[3] F Gao, S X Zhao, X S Li, and Y N Wang, 2009 *Phys. Plasmas* **16** 113502.

[4] X M Han, X L Wei, H J Xu, W Y Zhang, Y H Li, and Z X Yang, 2019 *Vacuum* **168** 108821.

[5] M A Lieberman and A J Lichtenberg, *Principles of Plasma Discharges and Materials Processing*, 2nd ed. (Wiley-Interscience, New York, 2005).

[6] A J Lichtenberg, V Vahedi, M A Lieberman and T Rognlien, 1994 *J. Appl. Phys.* **75** 2339.

[7] A J Lichtenberg, I G Kouznetsov, Y T Lee, M A Lieberman, I D Kaganovich, and L D Tsendin, 1997 *Plasma Sources Sci. Technol.* **6** 437.

[8] P Chabert and N Braithwaite, *Physics of Radio-Frequency Plasmas* (New York: Cambridge University press, 2011) p.305.

[9] M Lampe, W M Manheimer, R F Fernsler, S P Slinker and G Joyce, 2004 *Plasma Sources Sci. Technol.* **13** 15

[10] V I Kolobov and D J Economou, 1998 *Appl. Phys. Lett.* **72** 656.

[11] I G Kouznetsov, A J Lichtenberg and M A Lieberman, 1996 *Plasma Sources Sci. Technol.* **5** 662.

[12] I G Kouznetsov, A J Lichtenberg and M A Lieberman, 1999 *J. Appl. Phys.* **86** 4142

[13] S V Berezhnoj, C B Shin, U Buddemeier, and I Kaganovich, 2000 *Appl. Phys. Lett.* **77** 800.

[14] D Vender, W W Stoffels, E Stoffels, G M W Kroesen and F J de Hoog, 1995 *Phys. Rev. E* **51** 2436.

[15] K Kaga, T Kimura, T Imaeda and K Ohe, 2001 *Jpn. J. Appl. Phys.* **40** 6115.

[16] S X Zhao and A Q Tang, "*Study of transport process and discharge structure of inductively coupled electronegative plasmas via fluid model and analytic theory collaboration*", submitted to Physics

of Plasmas. Available: http://arxiv.org/abs/2111.01964
[17] S X Zhao and A Q Tang, 2021 *Chin. Phys. B* **30** 055201.
[18] https://fr.lxcat.net/instructions/.
[19] E G Thorsteinsson and J T Gudmundsson, 2010 *Plasma Sources Sci. Technol.* **19** 015001.
[20] S Tinck, W Boullart, A Bogaerts, 2008 *J. Phys. D: Appl. Phys.* **41** 065207.
[21] M V Malyshev and V M Donnelly, 2000 *J. Appl. Phys.* **88** 6207.